\documentclass[openany]{aa}
\usepackage[varg]{txfonts}
\usepackage{graphicx}
\usepackage{color}
\usepackage{float}
\graphicspath{ {./plots/} }
\usepackage{natbib}
\bibpunct{(}{)}{;}{a}{}{,}
\usepackage{multicol} 
\usepackage{array,multirow}
\usepackage{footnote}
\usepackage{url}
\usepackage[titletoc]{appendix}
\usepackage{tabularx}
\usepackage{soul}
\usepackage{babel,blindtext}
\usepackage{xcolor}
\usepackage{hyperref}
\hypersetup{colorlinks, allcolors = {blue!50!black}}
\usepackage{cleveref}
\usepackage{subfigure}
\usepackage{orcidlink}
\begin{document} 


\title{Lost in the curve: Investigating the disappearing knots in the blazar 3C 454.3}


\author{
Efthalia Traianou\inst{\ref{iaa},\ref{mpifr}} \orcidlink{0000-0002-1209-6500} \and 
Thomas P. Krichbaum\inst{\ref{mpifr}}\orcidlink{0000-0002-4892-9586} \and
Jos\'e L. G\'omez\inst{\ref{iaa}}\orcidlink{0000-0003-4190-7613} \and
Rocco Lico\inst{\ref{inaf},\ref{iaa}}\orcidlink{0000-0001-7361-2460} \and
Georgios Filippos Paraschos\inst{\ref{mpifr}}\orcidlink{0000-0001-6757-3098} \and
Ilje Cho\inst{\ref{iaa}}\orcidlink{0000-0001-6083-7521} \and
Eduardo Ros\inst{\ref{mpifr}}\orcidlink{0000-0001-9503-4892} 
Guang-Yao Zhao\inst{\ref{iaa}}\orcidlink{0000-0002-4417-1659} \and
Ioannis Liodakis\inst{\ref{finca}}\orcidlink{0000-0001-9200-4006} \and
Rohan Dahale\inst{\ref{iaa}}\orcidlink{0000-0001-6982-9034} \and
Teresa Toscano\inst{\ref{iaa}}\orcidlink{0000-0003-3658-7862} \and
Antonio Fuentes \inst{\ref{iaa}}\orcidlink{0000-0002-8773-4933} \and
Marianna Foschi\inst{\ref{iaa}}\orcidlink{0000-0001-8147-4993} \and
Carolina Casadio\inst{\ref{forth},\ref{crete}}\orcidlink{0000-0003-1117-2863} \and
Nicholas MacDonald\inst{\ref{mpifr},\ref{um}}\orcidlink{0000-0002-6684-8691} \and
Jae-Young Kim\inst{\ref{knu},\ref{mpifr}}\orcidlink{0000-0001-8229-7183} \and
Olivier Hervet \inst{\ref{sc}}\orcidlink{0000-0003-3878-1677} \and
Svetlana Jorstad\inst{\ref{bu},\ref{ais}} \orcidlink{0000-0001-6158-1708} \and
Andrei P. Lobanov\inst{\ref{mpifr}}\orcidlink{0000-0003-1622-1484} \and
Jeffrey~Hodgson\inst{\ref{sejong}}\orcidlink{0000-0001-6094-9291} \and
Ioannis Myserlis\inst{\ref{iram},\ref{mpifr}}\orcidlink{0000-0003-3025-9497} \and
Ivan~Agudo\inst{\ref{iaa}}\orcidlink{0000-0002-3777-6182} \and
Anton~J.~Zensus\inst{\ref{mpifr}}\orcidlink{0000-0001-7470-3321} \and
Alan P. Marscher \inst{\ref{bu}}\orcidlink{0000-0001-7396-3332} }

\institute{Instituto de Astrof\'\i sica de Andaluc\' \i a (IAA-CSIC), Glorieta de la Astronom\' \i a s/n, 18008 Granada, Spain \label{iaa} \\
\email{traianou@iaa.es}
\and Max-Planck-Institut f\"ur Radioastronomie, Auf dem H\"ugel 69, D-53121, Bonn, Germany \label{mpifr}
\and INAF $-$ Istituto di Radioastronomia, via Gobetti 101, 40129 Bologna, Italy \label{inaf}
\and Institute for Astrophysical Research, Boston University, 725 Commonwealth Avenue, Boston, MA 02215, USA \label{bu}
\and Saint Petersburg State University, 7/9 Universitetskaya nab., St. Petersburg, 199034 Russia \label{ais}
\and Department of Astronomy and Atmospheric Sciences, Kyungpook National University, Daegu 702-701, Republic of Korea \label{knu}
\and Department of Physics and Astronomy, Sejong University, 209 Neungdong-ro, Gwangjin-gu, 05006, Seoul, Republic of Korea \label{sejong}
\and Institute of Astrophysics, Foundation for Research and Technology $-$ Hellas, Voutes, 7110 Heraklion, Greece \label{forth}
\and Department of Physics, University of Crete, 71003 Heraklion, Greece \label{crete}
\and Institut de Radioastronomie Millim\'etrique, Avenida Divina Pastora, 7, Local 20, E18012 Granada, Spain \label{iram}
\and Santa Cruz Institute for Particle Physics and Department of Physics, University of California, Santa Cruz, CA 95064, USA \label{sc}
\and Finnish Centre for Astronomy with ESO, 20014 University of Turku, Finland \label{finca}
\and Department of Physics and Astronomy, The University of Mississippi, University, Mississippi 38677, USA \label{um}
            }

   \date{Version: \today; Received ???????; accepted ?????}


\abstract{One of the most well-known extragalactic sources in the sky, quasar \object{3C\,454.3}, shows a curved parsec-scale jet that has been exhaustively monitored with very-long-baseline interferometry (VLBI) over the recent years. In this work, we present a comprehensive analysis of four years of high-frequency VLBI observations at 43\,GHz and 86\,GHz, between 2013-2017, in total intensity and linear polarization. The images obtained from these observations enabled us to study the jet structure and the magnetic field topology of the source on spatial scales down to 4.6\,parsec in projected distance. The kinematic analysis reveals the abrupt vanishing of at least four new superluminal jet features in a characteristic jet region (i.e., region C), which is located at an approximate distance of 0.6\,milliarcseconds  from the VLBI core. Our results support a model in which the jet bends, directing the relativistic plasma flow almost perfectly toward our line of sight, co-spatially with the region where components appear to stop.}


\titlerunning{Lost in the Curve: Investigating the Disappearing Knots in the Blazar 3C 454.3}
\maketitle

\section{Introduction}
\label{sec:intro}

Supermassive black holes (SMBHs) are among the most intriguing and powerful objects in the universe. When gas and dust are trapped in their gravitational potential, tremendous amounts of energy are released, leading them to outshine their entire host galaxy \citep[e.g.,][]{1964SPhD....9..195Z}. Certain systems may even launch highly collimated, bipolar plasma jets, propagating over hundreds of kiloparsecs (kpc). When the jet axis of such an object is oriented at a small angle towards the observer's line of sight, we name it a blazar and it shows extreme variability across the electromagnetic spectrum \citep{1995PASP..107..803U}.

\indent The compact radio source \object{3C\,454.3} is a well-studied blazar in the Pegasus constellation, situated at $z=0.859$ \citep{1991MNRAS.250..414J}.
The SMBH at its center has an estimated mass between 0.5 and 1.5 $\times$ 10$^{9}$\,M$_\odot$ \citep{2002ApJ...581L...5W,2006ApJ...637..669L,2012MNRAS.421.1764S} and due to its striking broadband variability pattern is also known as the "Crazy Diamond" \citep[e.g.,][]{2009AIPC.1112..121V,2011MNRAS.410..368B}. Among the best techniques to study such an object as \object{3C\,454.3} is very long baseline interferometry.
VLBI provides high-resolution images of jet fine structures, probing regions down to parsec-scale distances from the central engine \citep[for a review see][]{2017A&ARv..25....4B}. Long-term VLBI monitoring observations have revealed a broad ensemble of propagating jet features, the nature of which has been interpreted as shocks \citep{1985ApJ...298..114M,2013A&A...551A..32F,2019arXiv191212358B}, flux enhancements moving in helical trajectories \citep{1965Natur.207..738R}, magneto-hydrodynamic instabilities \citep{2015ApJ...809...38M,2023NatAs...7.1359F}, or regions of magnetic reconnection \citep{2013MNRAS.431..355G,2017SSRv..207..291B}.
\\
\indent One of the earliest VLBI studies of the \object{3C\,454.3} \citep{1987Natur.328..778P}, unveiled structural changes in its jet that deviate from the theoretical expectations or observations of similar sources. They also reported the detection of a stationary feature at a projected radial distance of about 0.6\,mas ($\sim$ 4.6\,pc linear distance) from the innermost jet component. The nature of this knot was thoroughly investigated in subsequent studies \citep[e.g.,][and references therein]{Gomez_1999,2013ApJ...773..147J,2017ApJ...846...98J}, as it displayed complex kinematics and polarization patterns. According to \cite{1996MNRAS.278..861C}, this feature exhibits a dominant transverse magnetic field component, similar to a relativistically moving shock. However, other studies reported a jet break at these spatial scales, accompanied by highly polarized emission \citep{1996IAUS..175...33K}.
\\
\indent Between 2008-2010, the \object{Crazy Diamond} incited an extraordinary series of high-energy flares, achieving the highest $\gamma$-ray flux ever recorded from a non-transient source in the sky \citep{Ackermann_2010,2011ApJ...733L..26A,2011ApJ...736L..38V, 2012ApJ...758...72W}. Recent reports by \cite{Sarkar2021} and \cite{Liodakis_2020} highlighted the existence of a helically moving enhanced-emission region within a curved jet, leading to quasi-periodic oscillations in the high-energy regime and polarization fluctuations during 2013-2014. Other kinematic studies of the \object{3C\,454.3} jet revealed the existence of an arc-like structure at a radial distance from the VLBI core of about 2\,mas \citep{2013A&A...557A..37B, zamaninasab}. The unexpected detection of this structure was explained later on by multifrequency polarimetric VLBI imaging and magnetohydrodynamic simulations as being part of a large-scale ordered helical magnetic field, which was illuminated by a sudden energy injection from the jet base.
\\
\indent In this paper, we present a comprehensive analysis of the peculiar kinematics and magnetic fields topology of \object{3C\,454.3} from 2013 to 2017, focusing on the kinematics of the inner jet and the stationary feature located at 0.6\,milliarcsecond (mas) from the VLBI core. All calculations have been conducted using the following cosmological parameters: $\Omega_\mathrm{M}=0.27$, $\Omega_{\Lambda}=0.73$, $H_{0}=71$kms$^{-1}$Mpc$^{-1}$ \citep{2009ApJS..180..330K}. These parameters yield an angular scale of 7.70\,pc mas$^{-1}$ for \object{3C\,454.3}.

\section{Observations and data analysis}
\label{sec:obs_im}

\subsection{VLBA 43\,GHz data }

The 43\,GHz data that have been used in this work (notes on page 2), were obtained with the Very Long Baseline Array (VLBA), spanning the period from 2013 to 2017, within the VLBA-BU-BLAZAR program\footnote{\url{https://www.bu.edu/blazars/BEAM-ME.html}}. The program conducts regular monthly observations of a sample of $\gamma$-ray bright active galactic nuclei (AGN). A detailed description of the observations and data reduction can be found in \cite{2017ApJ...846...98J} and \cite{2022ApJS..260...12W}. For this study, we generated a total of 24 images observed within two months of the 86\,GHz data acquisition, of which only 8 are presented here.

\subsection{GMVA 86\,GHz data}
Eight epochs of \object{3C\,454.3} at 86\,GHz were obtained with the Global Millimeter VLBI Array (GMVA) within the program VLBI Images of Selected Gamma-ray Bright Blazars\footnote{\url{https://www.bu.edu/blazars/vlbi3mm}}, which is a supplement to the 43,GHz monitoring program. The data calibration was performed in accordance with the standard procedure for high-frequency VLBI data reduction for both data amplitudes and phases \citep[e.g.,][]{2012A&A...542A.107M}, via the Astronomical Image Processing System \citep[AIPS,][]{1990apaa.conf..125G}. For the amplitude calibration, we utilized measurements of the system temperature, gain curve, and atmospheric opacity of each telescope, to appropriately scale the measured visibilities. The phase calibration involved the removal of residual systematic delays, phase offsets, and time-dependent clock drifts in the data. The removal of the instrumental polarization leakage from the data, also known as D-terms \citep{1994cers.conf..207L}, was performed via the task LPCAL in AIPS, following the same process outlined in \cite{2019A&A...622A.158C}. Specifically, this technique involves the determination of station D-terms from a number of different sources that were observed during the same GMVA session. We exclude sources for which the parallactic angle coverage was less than 30$^\circ$. Subsequently, and for each telescope, we averaged the D-terms of the remaining sources and then applied them to the target source. The usage of the average D-terms results in higher dynamic ranges in polarization. The D-term magnitudes we obtained in this work are in the range of 1\%-15\%, consistent with previous studies \citep[e.g.,][]{2012A&A...542A.107M,2019A&A...622A.196K}. The uncertainties of D-terms computed using the relation $\sigma_\mathrm{m,D}= \sigma_\mathrm{D}\left(N_\mathrm{a} N_\mathrm{IF} N_\mathrm{s} \right)^{-1/2}$\citep{1994ApJ...427..718R}, where $\sigma_\mathrm{D}$ represents the standard deviation associated with the weighted average of D-term measurements (on the order of 1–4\%), $N_\mathrm{a}$ is the number of antennas, $N_\mathrm{IF}$ the number of Intermediate Frequencies (IFs), equal to 8 in our data sets, and $N_\mathrm{s}$ is the number of scans with independent parallactic angles. The noise level for each total intensity image was derived by averaging the root mean square (rms) values from multiple noise-only regions, using the IMSTAT function in \texttt{Difmap}. For the polarization images, we obtained the rms level of the same "empty" regions in Q and U images and then added them in quadrature. Finally, we performed the electric vector position angle (EVPA) absolute calibration by leveraging 3\,mm single-dish measurements from the IRAM 30-m antenna as part of the POLAMI program\footnote{See \cite{10.1093/mnras/stx2435,10.1093/mnras/stx2437} and http://polami.iaa.es}. A summary of the 43\,GHz and 86\,GHz images is given in \autoref{table:data43} and \autoref{table:data86}. 

\begin{figure*}[!htbp]
\centering
\includegraphics[width=1.0\linewidth]{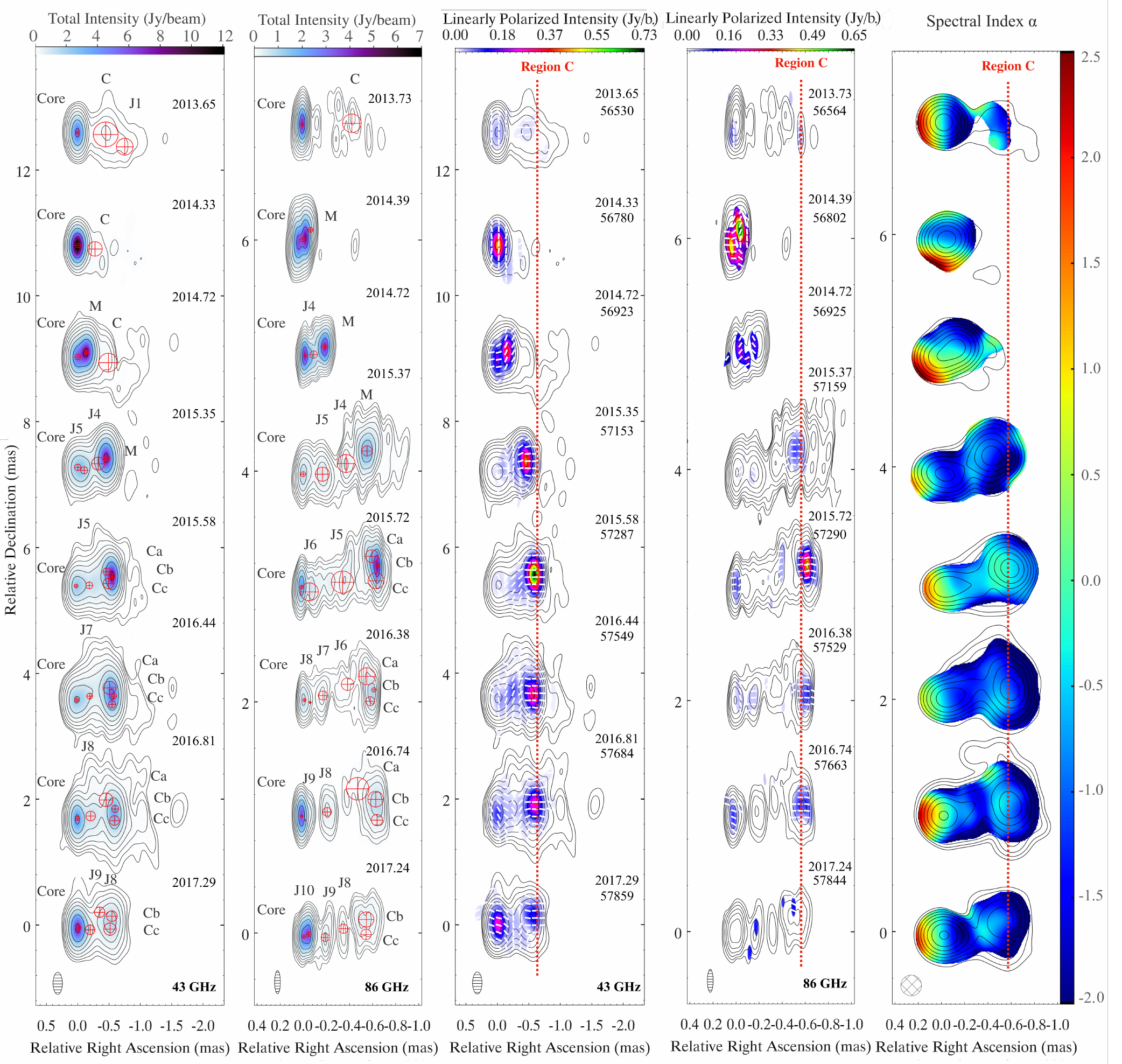}
\caption{Close-in-time total intensity, polarization, and spectral index images of the blazar \object{3C,454.3} at 43 and 86\,GHz from 2013 to 2017. The contour levels at 43,GHz are set to 0.11, 0.24, 0.57, 1.32, 3.07, 7.15, 16.33, 38.69, and 90\% of the peak total intensity of 12.2 Jy/beam; whereas at 86\,GHz they are set to 0.49, 1.17, 2.80, 6.66, 15.87, 37.79, and 90\% of the peak total intensity of 4.7 Jy/beam. The 43\,GHz images are convolved with an indicative common restoring beam of $0.35\times0.15$\,mas oriented at position angle (PA) $0^\mathrm{o}$, whereas the 86\,GHz maps are convolved with a beam of $0.2\times0.05$\,mas. \textbf{Left two panels}: Stokes I images of the source. The data were imaged using a uniform weighting scheme, and the color scale visualizes the total intensity of each image. The red circles represent all the 2D Gaussian components that model the flux density distribution along the jet at each epoch. \textbf{Middle two panels}: polarization images of the same source. The color scale visualizes the linearly polarized intensity of each image, whereas the white sticks show the EVPAs. \textbf{Right panel}: Spectral index maps between the 43\,GHz and 86\,GHz. These maps were obtained after convolution with a mean circular beam of 0.16\,mas and a mean pixel size of 0.007\,mas. Lastly, the vertical red dotted line indicates the approximate location of region C.}
\label{fig:vlbi_maps}
\end{figure*}

\subsection{VLBI imaging and model-fitting}
\label{sec:models}

The fully calibrated data from all epochs were imported into \texttt{Difmap}, an interactive program for synthesis imaging \citep{1997ASPC..125...77S}. For the GMVA data, we coherently time-averaged them to 10 sec, following the method described by \cite{1997ASPC..125...77S}. Then, by employing the CLEAN algorithm \citep{1974A&AS...15..417H} and {\tt SELFCAL} procedure, we generated images of the source across all epochs and frequencies, both in terms of the total intensity and linear polarization.
The results are presented in \autoref{fig:vlbi_maps} and described in \autoref{table:data43} and \autoref{table:data86}.
Prior to imaging, we conducted a thorough investigation of the visibilities, with a particular focus on the high-frequency 86,GHz data, flagging any erroneous data points and noticeable outliers.
Once the final CLEAN images were obtained, we employed the {\tt MODELFIT} algorithm in \texttt{Difmap} to parameterize the jet brightness distribution by fitting two-dimensional (2D) circular Gaussian components to the self-calibrated data. 
Each component was cross-identified between different epochs by comparing the flux, radial separation with respect to the core, and position angle parameters \citep[see also][]{2022A&A...665A...1P}.
The uncertainties associated with each parameter of the Gaussian components were formally evaluated based on the local signal-to-noise ratio (S/N) in the image surrounding each component \citep{1999ASPC..180..301F,2005astro.ph..3225L,2012A&A...537A..70S}. However, for the flux density uncertainty, we adhered to more conservative criteria, as the formal errors obtained via this method appeared too small. Therefore, we set the flux density uncertainty to 10\% of the measured value, as suggested by \cite{2009AJ....138.1874L}. All parameters of the fitted Gaussian components are provided in \autoref{table:knots43a} and \autoref{table:knots86}.

\begin{table*}
\centering
\caption{Image parameters of the presented 43\,GHz data. }
\begin{tabular}{@{}ccccccccccccc@{}}
\hline \hline
 Epoch   & $b_\mathrm{maj}$ 
                & $ b_\mathrm{min}$ 
                      & PA      & S$_\mathrm{peak}$ 
                                        & S$_\mathrm{total}$ 
                                                & S$_\mathrm{rms}$ 
                                                     & P$_\mathrm{tot}$ 
                                                            & P$_{rms}$  
                                                                & $m$    & $\Delta m$ 
                                                                              & $\chi$ 
                                                                                    & $\Delta \chi$ 
                                                                              \\
 (Years) & (mas)
               & (mas)& ($^{\circ}$)   & (Jy/b)& (Jy) & (mJy/b)       
                                                     & (Jy) & (mJy/b) 
                                                                & (\%) & (\%) & ($^{\circ}$)
                                                                & ($^{\circ}$)
                                                                \\
                                                                (1) & (2) & (3) & (4) & (5) & (6) & (7) & (8) & (9) & (10) & (11) & (12)
                                                                & (13)\\
\hline 
2013.65 & 0.32 & 0.15 & $-10$ & 5.33  & 6.40  & 1  & 0.03 & 2 & 0.5 & 0.2 & 1 & 11 \\
2014.33 & 0.32 & 0.13 & $-3$ & 12.10 & 14.96 & 3  & 0.56 & 4 & 3.7 & 1.2  & $-$5 & 10 \\
2014.72 & 0.37 & 0.13 & $-3$ & 8.13  & 15.60 & 1  & 0.36 & 8 & 2.3 & 0.7  &  $-$24 & 10 \\
2015.35 & 0.33 & 0.14 & 12   & 7.95  & 15.50 & 2  & 0.40 & 4 & 2.6 & 0.8  & $-$85 & 11 \\
2015.58 & 0.34 & 0.14 & $-9$ & 8.15  & 16.20 & 2  & 0.82 & 3 & 5.1 & 1.6  &  $-$87 & 10 \\
2016.44 & 0.35 & 0.14 & $-5$ & 5.19  & 13.90 & 1  & 0.32 & 2 & 2.3 & 0.7  & $-$95 & 10 \\
2016.81 & 0.33 & 0.14 & $-4$ & 4.10  & 13.80 & 1  & 0.34 & 3 & 2.5 & 0.8  & $-$100 & 11 \\
2017.29 & 0.32 & 0.12 & $-5$ & 7.12  & 12.86 & 1  & 0.42 & 3 & 3.3 & 1.0  & $-$105 & 10 \\\hline 
\end{tabular}
\tablefoot{Columns from left to right: (1) observed epoch, (2) beam major axis, (3) beam minor axis, (4) beam position angle, (5) peak intensity, (6) total flux density, (7) I rms level, (8) total polarized flux density, (9) P rms level, (10) integrated fractional polarization, (11) uncertainty on fractional polarization, (12) electrical vector position angle, and (13) uncertainty on the electrical vector position angle.}
\label{table:data43}
\end{table*}

\begin{table*}
\centering
\caption{Image parameters at 86\,GHz data. }
\begin{tabular}{@{}ccc ccc cc ccc ccc@{}}
\hline\hline
\noalign{\smallskip}
Epoch   & $b_\mathrm{maj}$  & $ b_\mathrm{min}$ & PA  & S$_\mathrm{peak}$ & S$_\mathrm{total}$ & S$_\mathrm{rms}$ & N$_{\rm{ant}}$ &  N$_{\rm{scan}}$ & P$_\mathrm{tot}$ & P$_\mathrm{rms}$ & $m$ & $\Delta$$m$ & $\chi$ \\ 
(Years) & (mas)& (mas)  & ($^{\circ}$)    & (Jy/b)   & (Jy)   & (mJy/b) &     &   & (Jy) & (mJy/b) & (\%)& (\%)& ($^{\circ}$) \\
(1) & (2) & (3) & (4) & (5) & (6) & (7) & (8) & (9) & (10) & (11) & (12) & (13) & (14) \\
\noalign{\smallskip}
\hline 
\noalign{\smallskip}
2013.73 & 0.28 & 0.05   & $-12$ & 4.45 & 5.44   & 1.8 & 6   & 11 & 0.13 & 10 & 2.4 & 0.7 & 25 \\
2014.39 & 0.28 & 0.04   & $-6$  & 4.70 & 11.33  & 1.9 & 8   & 13 & 0.21 & 20 & 1.9 & 0.6 & $-$33 \\
2014.72 & 0.28 & 0.05   & $-10$  & 4.78 & 14.53  & 3.0 & 10  & 12 & 0.05 & 20 & 0.3 & 0.2 & $-$36 \\
2015.37 & 0.28 & 0.05   & $-9$  & 1.75 & 7.16   & 0.3 & 13  & 15 & 0.17 & 10 & 2.4 & 0.7 & $-$94 \\
2015.72 & 0.28 & 0.05   & 15    & 6.80 & 12.95  & 2.5 & 10  & 14 & 0.79 & 10 & 6.1 & 1.9 & $-$79 \\
2016.38 & 0.23 & 0.05   & $-11$ & 1.28 & 4.19   & 1.0 & 10  & 9  & 0.16 & 10 & 3.9 & 1.2 & $-$89 \\
2016.74 & 0.21 & 0.05   & $-9$  & 3.79 & 8.38   & 4.0 & 13  & 5  & 0.42 & 10 & 5.0 & 1.6 & $-$101 \\
2017.24 & 0.28 & 0.05   & 1     & 4.05 & 9.08   & 4.0 & 13  & 7  & 0.28 & 30 & 3.1 & 1.0 & $-$96 \\
\noalign{\smallskip}
\hline 
\end{tabular}
\tablefoot{Columns from left to right: (1) observed epoch, (2) beam major axis, (3) beam minor axis, (4) beam position angle, (5) peak intensity, (6) total flux density, (7) I rms level, (8) participating antennas at the given session, (9) number of independent scans with different parallactic angles, (10) total polarized flux density, (11) P rms level, (12) integrated fractional polarization, (13) uncertainty on fractional polarization, and (14) electrical vector position angle. The uncertainties on EVPA are $\approx$5$^\circ$ for all epochs.}
\label{table:data86}
\end{table*}

\section{Results}
\label{sec:analysis_results}

\subsection{Jet structural evolution and kinematics}

\begin{figure*}[!t]
\centering
\includegraphics[scale=0.5]{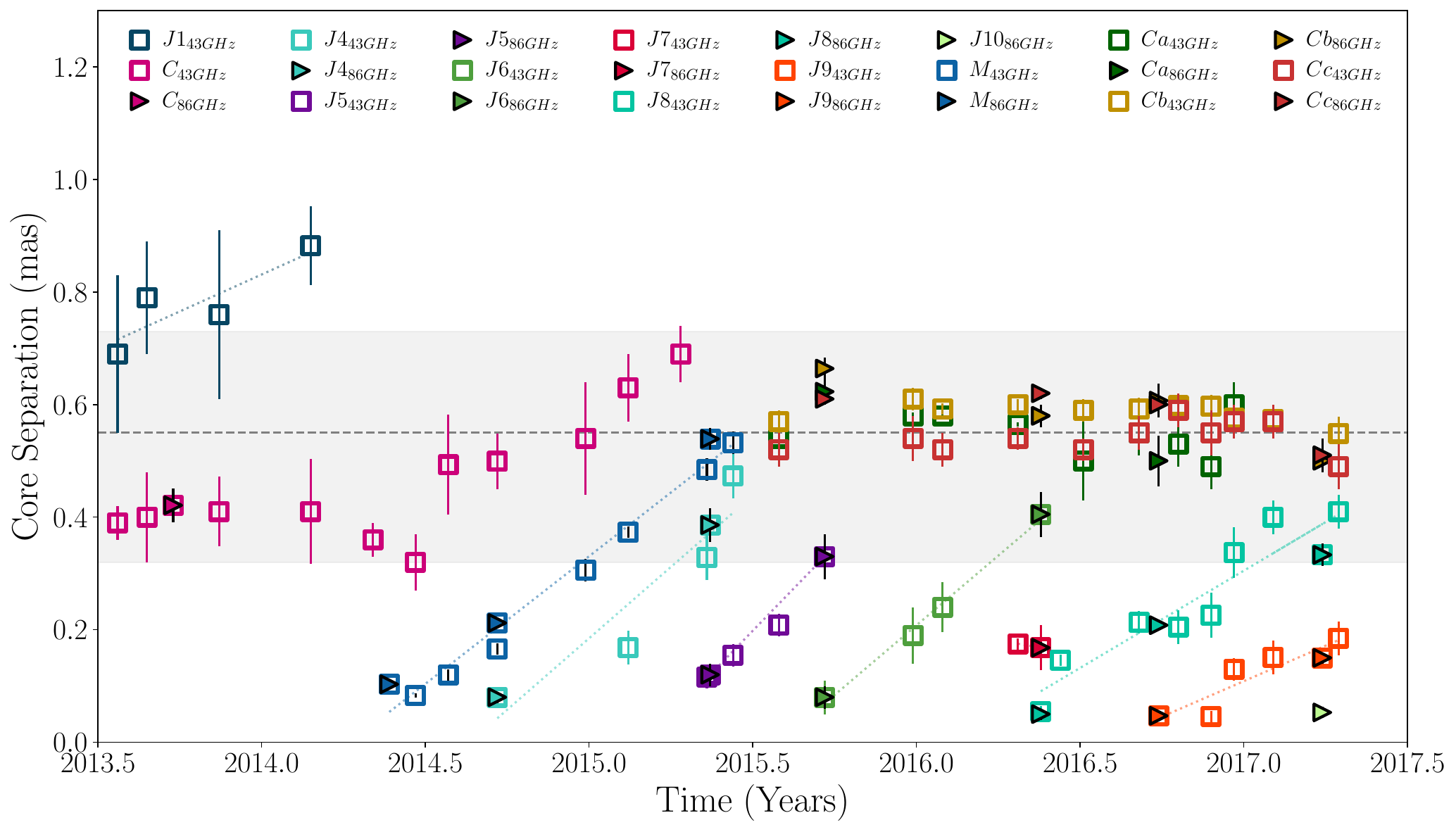}
\includegraphics[scale=0.25]{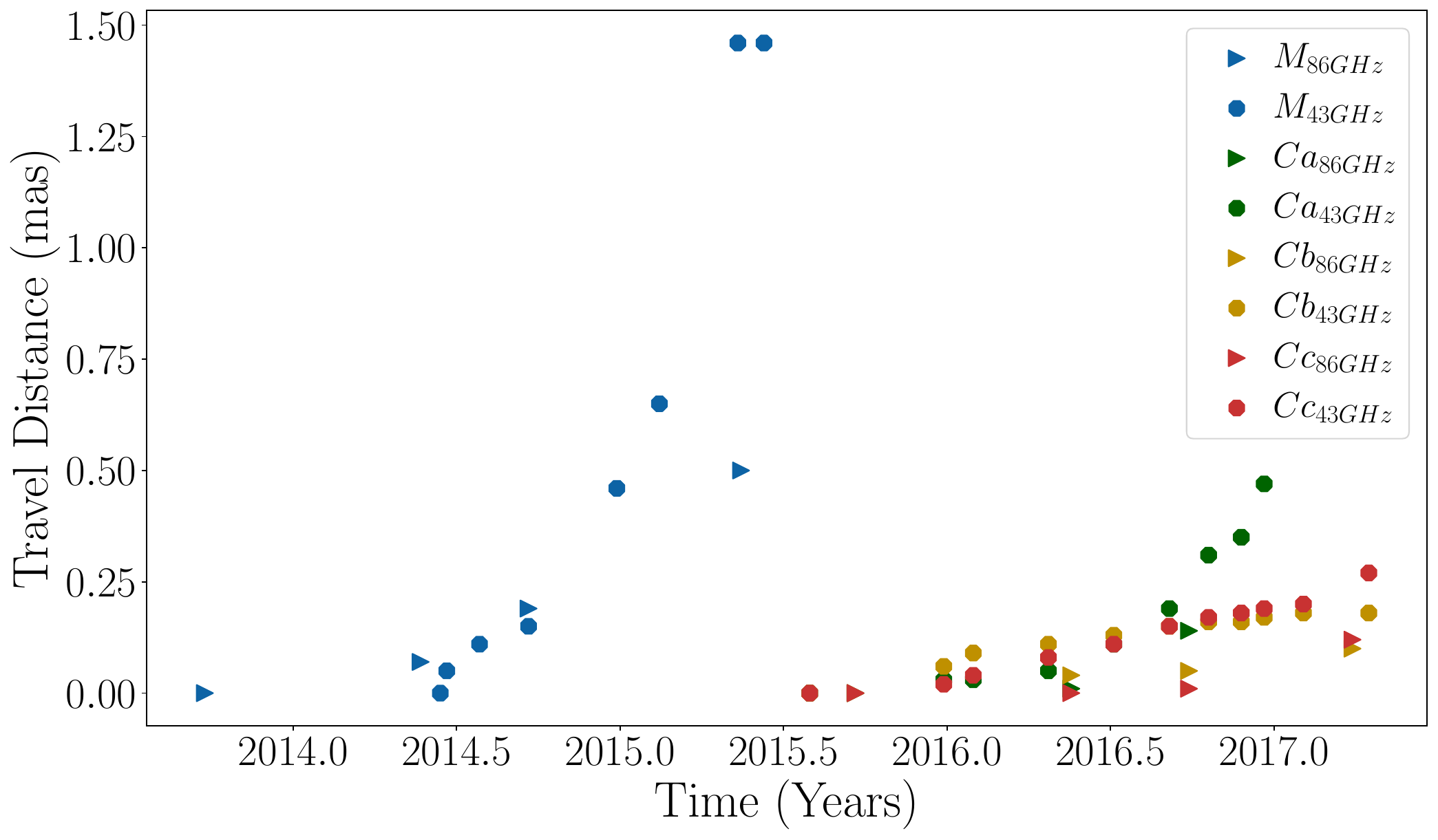}
\includegraphics[scale=0.135]{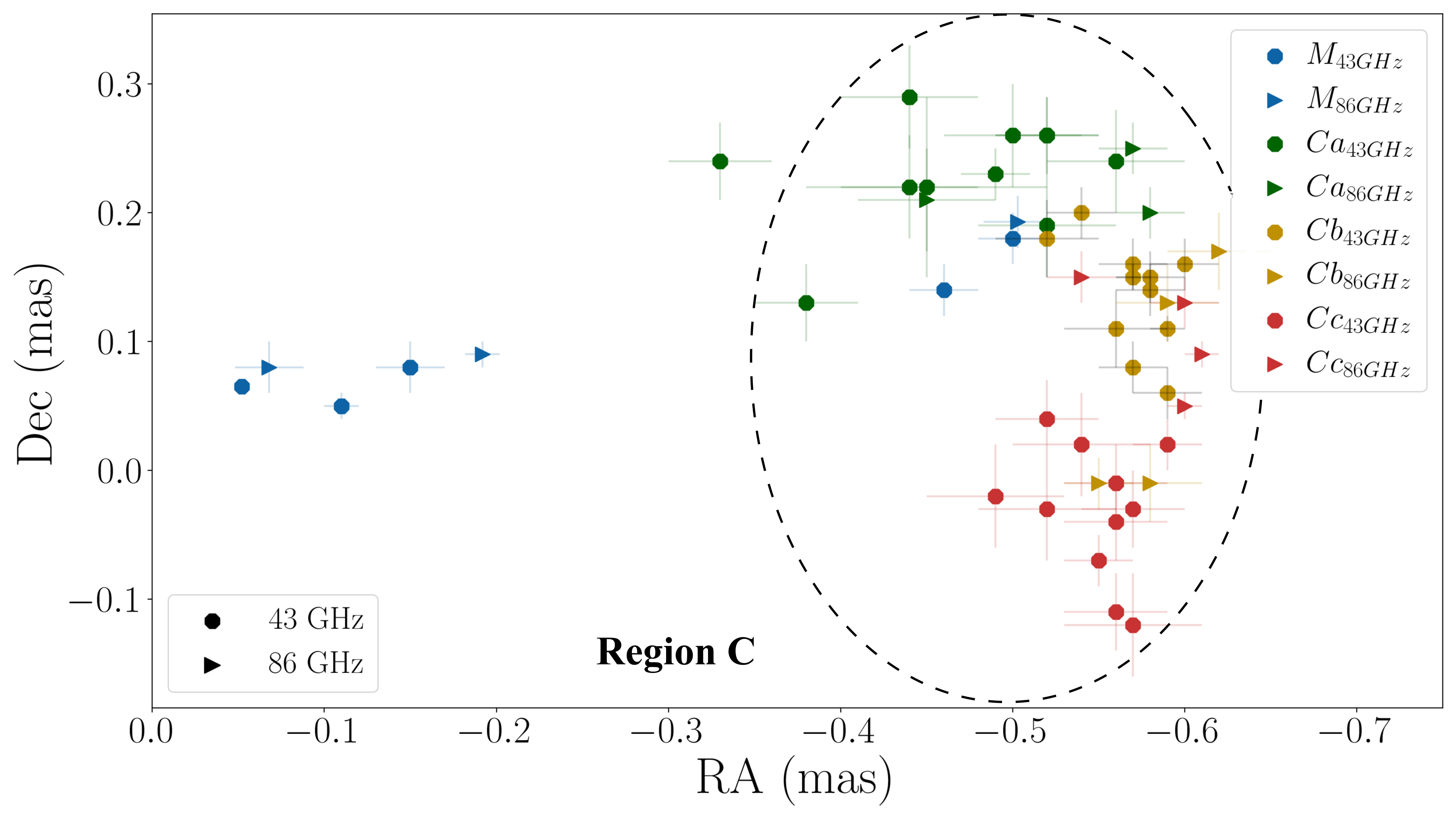}
\caption{\textbf{A comprehensive view of the spatial and temporal dynamics of \object{3C 454.3} during the period 2013-2017. Upper panel}: Radial core separation versus time of all the identified jet components in 3C\,454.3. The dashed, horizontal line designates the mean position of region C during our observing interval, whereas the gray shadowed area represents the positional range that region C has been reported by this and previous studies. \textbf{Lower-left panel}: Travel distance evolution of knots M, Ca, Cb, and Cc. \textbf{Lower-right panel}: Relative right ascension and declination with respect to the VLBI core of components M, Ca, Cb, and Cc. The dashed, grey ellipsis highlights the location of region C relative to the core.
}
\label{fig:kinematics}
\end{figure*}

As shown in \autoref{fig:vlbi_maps}, the imaging and Gaussian model fitting of \object{3C,454.3} at 43,GHz and 86,GHz revealed the presence of a complex and variable western-oriented jet structure. In September 2013 (56564 MJD), the source was clearly characterized by a core-dominated morphology with a continuous one-sided jet. For the purposes of this analysis, we identify the easternmost total intensity component as the VLBI core. Around early 2014 (56802 MJD), we detect the appearance of a new jet feature, named M, which subsequently reached a flux density level that surpassed the core emission by 250\%. The propagation of this knot along the jet seems to halt when reaching a radial distance of $\sim$0.6\,mas, designated here as "region C" (red vertical line in \autoref{fig:vlbi_maps}). At the same time with the arrival of M, the region begins to develop significant elongation in the north-south direction, extending by approximately 0.7 $-$ 0.8\,mas, modeled by a cluster of three knots,
which we label as Ca, Cb, and Cc. The jet sustained this morphology for roughly two years. The direction of this elongation relative to the bulk plasma flow is discussed thoroughly in \Cref{sec:discussion}.
\\
\indent Besides component M, we identified 8 additional moving features, labeled from J1, J4, J5, up through J10. The dynamics of the moving components was derived by following the method described in \cite{2020A&A...634A.112T}, and
for all knots following near-ballistic trajectories, we calculated the angular proper motion on the sky through a linear fit of their radial core separation as a function of time, up to region C. The apparent speed ($\beta_\mathrm{app}$) and the critical viewing angle ($\theta_\mathrm{c}$, the angle at which the Doppler boosting and the apparent speed are maximum) of each knot were computed by the relations \citep[e.g.,][Eq. 8]{1966Natur.211..468R,2016A&A...586A..60K}:
\begin{equation}
\label{eq:theta_c}
\centering
\beta_\mathrm{app}=\frac{\beta \sin \theta}{ 1-\beta \cos \theta} = \frac{\mu D_L}{c(1+z)}, ~~ \theta_\mathrm{c} = {\rm sin}^{-1} \frac{1}{\left( \beta_\mathrm{app}^{2}+1 \right)^{1/2} },
\end{equation}
\noindent where $\beta$ is the plasma velocity, $\mu$ is the proper motion in rad~s$^{-1}$, $D_{L}$ is the luminosity distance in m, $z$ is the source redshift, and $c$ is the speed of light in m\,s$^{-1}$ \citep[e.g.,][]{1999astro.ph..5116H}.
\\
\indent In the upper panel of \autoref{fig:kinematics} we present the results of this analysis, which reveals that all the newly emerged knots upstream of M, appear also to halt upon reaching region C. In the two lower panels of \autoref{fig:kinematics}, we illustrate the relative coordinates (right ascension and declination) of all the modeled components. Interestingly, we notice that the travel distance of M becomes zero in region C, whereas Ca, Cb, and Cc, although they do not move radially (see \autoref{fig:kinematics}), they do exhibit curvilinear movement.
\\
\indent Kinematic analysis of the remaining moving centroids reveals notably large superluminal velocities (see \autoref{table:speeds}), ranging from 10 to 30\,$c$ at 43\,GHz and 86\,GHz. For these speeds, the critical viewing angle ranges between 5.8 and 2 degrees, respectively. A comparison of the $\beta_\mathrm{app}$ between 43\,GHz and 86\,GHz does not reveal a systematic difference between the two bands.
\\
\indent The broad range of apparent velocities estimated in the innermost region of the jet may indicate a bend away from our line of sight.
In extreme environments, even small differences in jet inclination can lead to substantial fluctuations in the apparent velocity.
In \cite{2021A&A...653A...7Q}, this phenomenology was interpreted as the result of the existence of a binary SMBH system in the center of \object{3C 454.3}, displaying a double precessing jet system. In this regime, two sets of superluminal knots are ejected in different directions, originating from two distinct jets precessing with a consistent period of 10.5 years. Other alternative scenarios involving component motion along spatially curved trajectories, intrinsic jet acceleration combined with regions of slower velocity; alternatively, shocks are also plausible.
In this work, we consider that the physical origin of the relatively large variation in component speeds along the jet remains unclear, and that future, more detailed kinematic studies are needed.
A comprehensive description of the modeling parameters of all knots is presented in \autoref{table:knots43a} and \autoref{table:knots86}, while the kinematic results are presented in \autoref{table:speeds}.
Finally, we point out that our findings are in good agreement with the values presented in \cite{2022ApJS..260...12W}, despite their utilization of a substantially larger number of epochs in their analysis.

\subsection{Polarization and magnetic field topology}

A defining observable of a blazar is the linear polarization of emission features in the jet, which at 43\,GHz typically ranges from a few to tens of percent \citep{Marscher2002,Jorstad2007}. The middle two panels of \autoref{fig:vlbi_maps} show the close-in-time polarization images of \object{3C 454.3} at 43 and 86\,GHz. The evolution of the polarized flux density at 43\,GHz reveals a highly variable degree of polarization, $m$, in the core region and further downstream in the jet (presented in \autoref{table:data43} and \autoref{table:data86}), whereas component M exhibits a prominent polarization signature after its first detection and until its arrival at region C. A topic of particular interest is the epoch 57153\,MJD (2015.35), as M exhibits more linearly polarized flux than the core, indicating that the magnetic field within the knot is highly ordered and perpendicularly aligned with the jet direction. After M arrives at region C, the entire region dominates the polarized emission of the jet. The EVPAs at the core swing between perpendicular and parallel to the jet axis, implying that the magnetic field oscillates within the jet. This could be caused by various physical processes, such as helical patterns and/or filaments in the jet, oblique shocks, or even the interaction of the jet with the ambient medium \citep[e.g.,][and references therein]{2013EPJWC..6106003K,Cohen_2018,2020A&A...636A..79C}. In region C, we detected highly ordered EVPAs; however, we cannot be sure about their relative orientation with respect to the jet flow, as in this region the jet direction is a matter of debate.

The distribution and evolution of the polarized flux density and EVPAs orientation at 86,GHz show an almost identical pattern with the 43,GHz, except for the epochs 2014.33 and 2014.39, which by visual inspection reveal a rotation of $\sim$90$^\circ$ between the two frequencies. A 90$^\circ$ swing in the EVPA could indicate a transition from optically thick plasma (at 43,GHz) to optically thin plasma (at 86,GHz). Another possible explanation is that the orientation of the magnetic field at the base of the jet varies due to the rotation of the jet around its own axis. Additionally, it should be noted that the beam sizes at 43\,GHz and 86\,GHz differ, and blending effects from a complex, multi-component polarized sub-structure may mimic the observed swing in EVPA.

These findings may indicate the presence of large-scale helical magnetic fields, illuminated by the propagating M \citep{2016Galax...4...32R,2016Galax...4...45J}, further supporting the findings of \cite{10.1093/mnras/stt1816}. However, the fact that we do not detect any significant change in the EVPAs orientation between frequencies indicates that, during our observing interval, the magnetic field is organized in such a way that it appears uniform on the spatial scales probed by the VLBI observations. \autoref{table:data43} and \autoref{table:data86} show the polarization properties of each image, whereas the uncertainties of the polarization parameters were calculated according to the process described in \cite{2014A&A...571A..54L}.

\subsection{Pixel-based spectral decomposition}
\label{sec:spix}

The next step in this analysis is to reconstruct the spectral index along the jet by using pairs of images at different frequencies. The spectral index, $\alpha$, is defined as $\alpha=\ln\left(S_{1}/S_{2}\right)/ \ln\left(\nu_{1}/\nu_{2}\right)$, where $S_{1,2}$ are the flux densities in each pixel, and $\nu_{1,2}$ are the frequencies of each image. In this analysis we use $S \sim \nu^{+\alpha}$. A critical consideration in generating spectral index maps is the selection of an appropriate common beam and pixel size. The selection of a too-small beam size will introduce image artifacts, whereas an overly large pixel size will lead to the loss of sensitivity. To mitigate these concerns, we employed a standard approach of utilizing a circular beam, its radius being the maximum typical resolution of the lower frequency data set \citep{2013A&A...557A.105F}. The final, common parameters for all pairs are b = 0.16\,mas and a pixel size of 0.007\,mas. No spectral index was calculated for pixels with a flux density smaller than five times the rms noise level. The alignment of the images was achieved using 2D cross-correlation analysis, focusing on a region within the optically thin segment of the jet. The resulting average shift across all epochs is on the order of $\Delta$RA $= 25 \pm 23\,\mu$as and $\Delta$Dec $= 37 \pm 19\,\mu$as.
It is noteworthy that the orientation of the spectral index gradient in the core region varies over time. Although this is most likely due to the uncertainties in the image shift, for the sake of completeness, we report that similar findings in numerically simulated images of Sgr~A$^{*}$ \citep{2013MNRAS.432.2252D} appeared when non-axisymmetric standing shocks from eccentric fluid orbits in misaligned accretion flows dominate the emission, influencing the spectral index and its variability. Also, spectral index maps of 3C\,84 from quasi-simultaneous observations at different frequencies revealed a time-variable orientation of the spectral index gradient, attributed to potentially helical, bend, or rotating trajectories of ejected features that could be aligned or misaligned with the line of sight \citep{2022A&A...665A...1P}. 
As we move further downstream in the jet, $\alpha$ decreases with distance, which is consistent behavior for blazar jets. Spectral indices much lower than -2 are likely due to beam resolution effects and therefore may not be real. The spectrum of region C is predominantly optically thin, spanning values from 0 to -1.5. A distinct spectral behavior emerges in late 2015, where the tri-component structure (Ca, Cb, and Cc) becomes visible in our images. The increased opacity observed in this region may be attributed to a jet bending, which (when illuminated) reveals a greater amount of material. In subsequent epochs, there is a gradual decline in the flux density within the whole region. We present the results in the right column of \autoref{fig:vlbi_maps}.

\subsection{Brightness temperature}

\begin{figure}[htbp]
    \centering
    \subfigure{\includegraphics[width=0.45\textwidth]{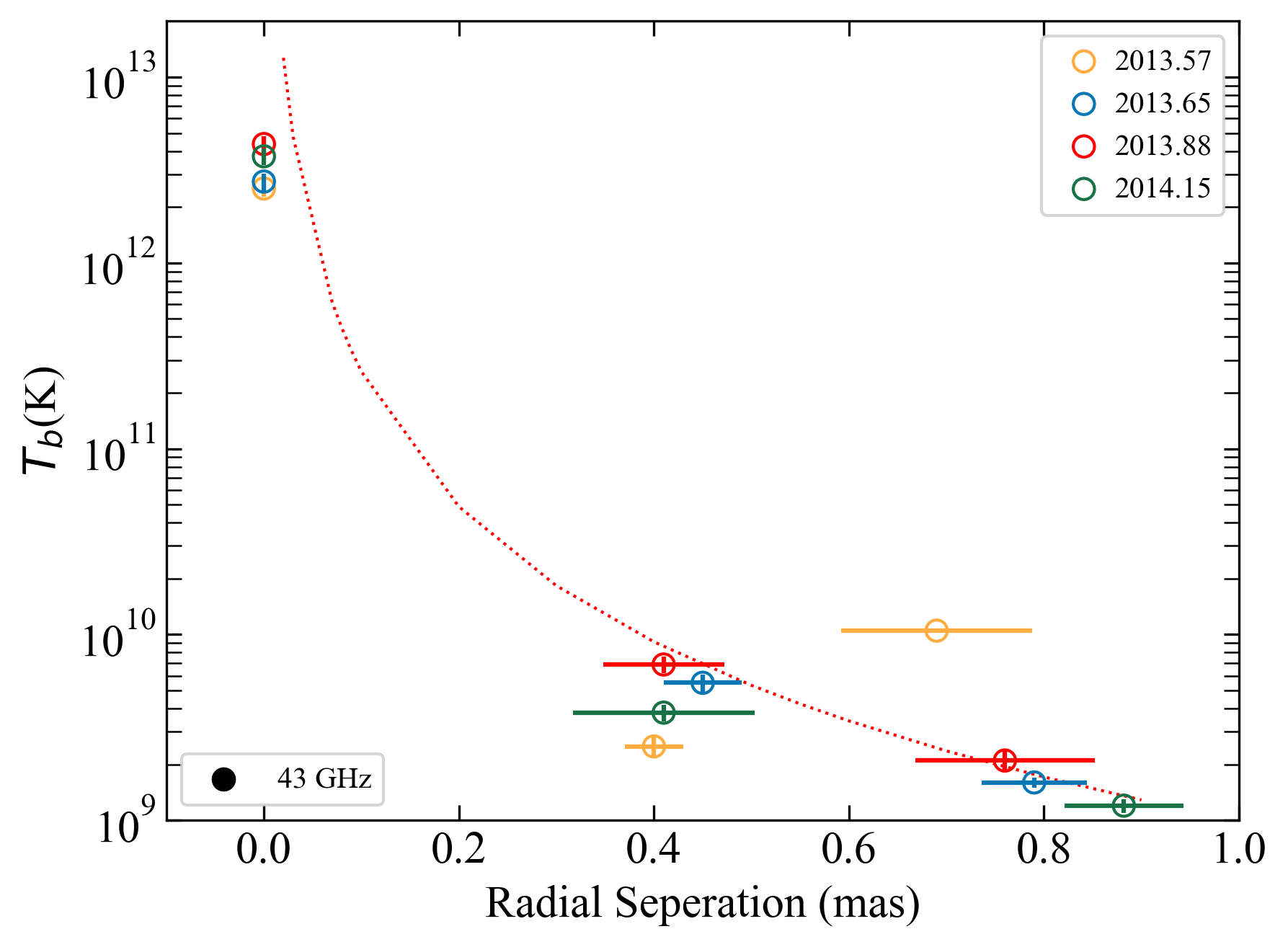}}
    \subfigure{\includegraphics[width=0.45\textwidth]{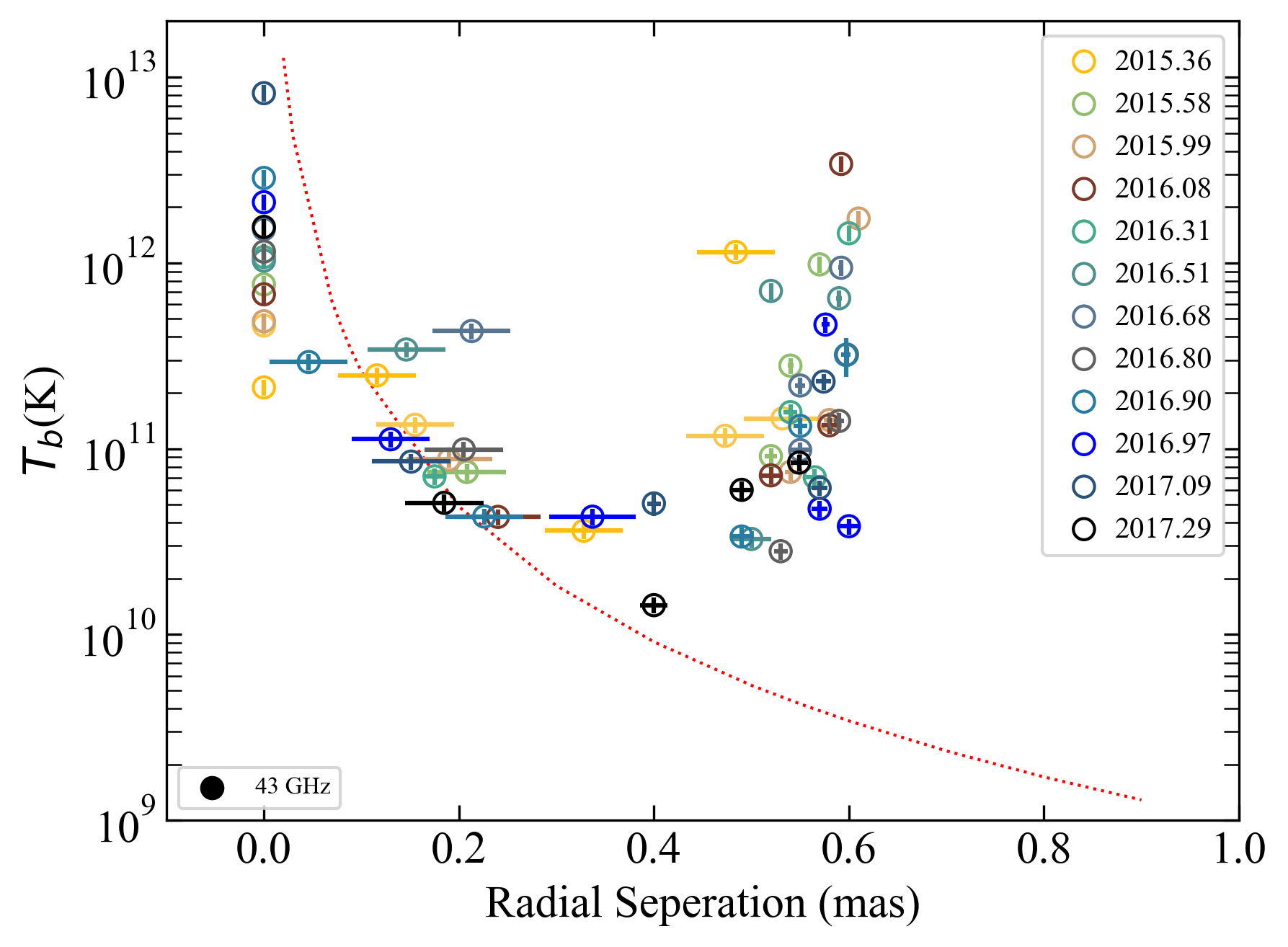}}
    \subfigure{\includegraphics[width=0.45\textwidth]{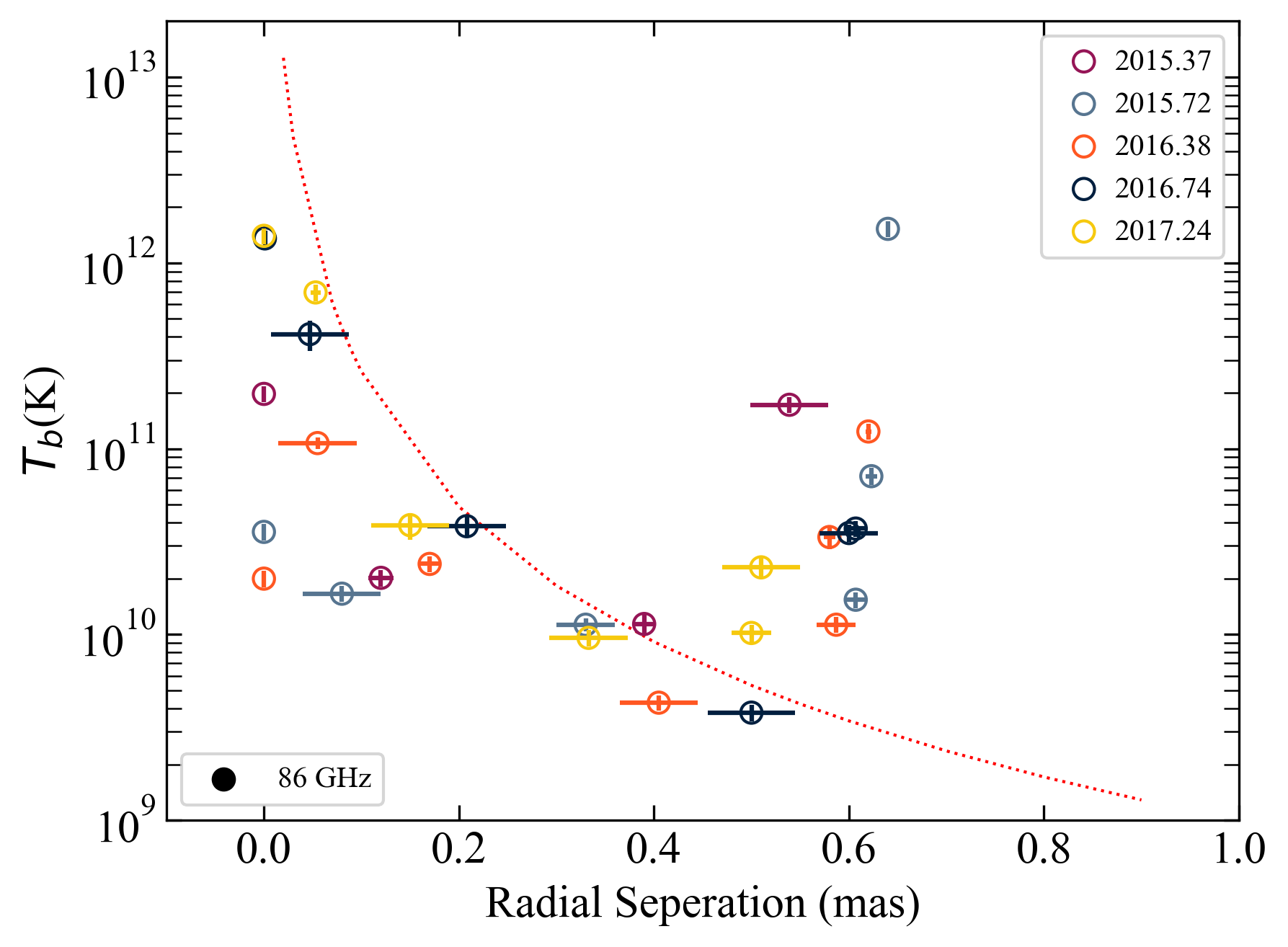}}
    \caption{Brightness temperature evolution of all the detected components in observer's frame at 43 and 86\,GHz. \textbf{Top}: $T_\mathrm{b,obs}$ evolution versus time, before the appearance of knot M at 43\,GHz. \textbf{Middle}: $T_\mathrm{b,obs}$ evolution versus time, after the appearance of the knot cluster Ca, Cb, Cc at 43\,GHz. \textbf{Bottom}: $T_\mathrm{b,obs}$ evolution versus time, after the appearance of the knot cluster Ca, Cb, Cc at 86\,GHz. In all three panels the red dotted line is the fit of the radial distribution of $T_\mathrm{b,obs}$ before the appearance of knot M.}
    \label{fig:tb}
\end{figure}

The energy density of plasma flow can be quantified by a representative temperature known as the brightness temperature, $T_\mathrm{b}$, which corresponds to the temperature of the source if it was radiating as a black body.  In this work, we have calculated the $T_\mathrm{b}$ for each VLBI knot in the observer's frame ($T_\mathrm{b,obs}$). This calculation is based on the fitted flux densities and sizes of the modelfit components, via the relation \citep[e.g.,][]{2019A&A...622A..92N}: 

\begin{equation}
    T_\mathrm{b,obs} = 1.22 \times 10^{12} \frac{S}{\theta_\mathrm{obs}^{2} \nu^{2}} \left( 1+z \right) ~~[K], 
    \label{eq:Tb}
\end{equation}

\noindent with $S$ denoting the component flux density in Jy, $\theta_\mathrm{obs}$ the apparent size of the emitting region in $\mathrm{mas}$, and $\nu$ the observing frequency in GHz. For unresolved modelfit components, we set $\theta_\mathrm{obs} = \theta_\mathrm{min}$ \citep{2005astro.ph..3225L} of each knot (6\% of all detected knots at 43\,GHz and 10\% of the 86\,GHz), and consider the resulting estimate of $T_\mathrm{b,obs}$ as a lower limit. 
The brightness temperature estimated for each knot is presented in the column 8 of \autoref{table:kinem43} and \autoref{table:kinem86}, and also displayed in \autoref{fig:tb} as a function of their radial distance from the core. 

At 43\,GHz, the core $T_\mathrm{b,obs}$ fluctuates and appears to be causally connected to the appearance of new knots. Its values consistently exceed the equipartition temperature limit, $T_\mathrm{b,eq}=5\times 10^{10}$ K \citep{1994ApJ...426...51R} and, in many cases, the inverse-Compton limit, $T_\mathrm{b,IC}\sim 5 \times 10^{11}$ K \citep{1969ApJ...155L..71K}, indicating emission amplification by relativistic beaming \citep{2010A&A...512A..24S,Kovalev_2016}. The remaining newly ejected knots have $T_\mathrm{b,obs}$ values around the equipartition limit and above, whereas the outermost knot, J1 ($\approx$0.8\,mas), exhibits values below the $T_\mathrm{b,eq}$, indicating possible magnetic dominance in this segment of the jet. Similar values were obtained for the knot "C," the precursor of region C. Knot C is of special interest as, even if considered as a stationary feature \citep[e.g.,][and references therafter]{1987Natur.328..778P,Gomez_1999,2001ApJS..134..181J}, it appears to change position, by $\sim$ 0.2 mas over $\sim$ 600 days (mid-2013 to early 2015). \cite{Liodakis_2020} explained this behavior as the core drifting downstream, during the passage of the superluminal knot K14 through the core region, combined with the increase of the opacity of the same region due to blending effects. 

Moving outward along the jet, the radial brightness temperature distribution of all components before the appearance of M was generally declining, following the power law $r^{-2.4 \pm 0.1}$ (top panel of \autoref{fig:tb}). Typically, the radial brightness temperature distribution in blazar jets shows such a decline, owing to the expansion of the jet and the adiabatic cooling of the emitting plasma \citep[e.g.,][]{kadler2005,2012A&A...544A..34P}. Nevertheless, after the arrival of M and the appearance of the knots cluster Ca, Cb, and Cc, the local brightness temperature increased to $\sim 170 \times 10^{10}$\,K, about 550 times higher than the expected value of $\sim 0.3 \times 10^{10}$\,K. We report remarkably high observed brightness temperatures ($T_\mathrm{b,obs}$) for M as well, reaching the highest value recorded within our observation period (in 2014.47 $\approx9 \times 10^{12}$ K). 

The limited number of 86\,GHz epochs, didn't allow us to study the evolution of the radial brightness temperature distribution prior to ejection of M. However, a comparison between 43\,GHz and 86\,GHz values revealed a decreasing trend in brightness temperature with frequency, as expected, since the radiative losses are more efficient at higher frequencies \citep{1962SvA.....6..317K}. It is important to note that these values are influenced by the Doppler factor, which depends on the viewing angle of the jet. Specifically, the observed $T_\mathrm{b,obs}$ is related to the source frame brightness temperature ($T_\mathrm{b,int}$) through the relation:

\begin{equation}
    T_\mathrm{b,int} = \frac{T_\mathrm{b,obs}}{\delta}  ~~[K],
\end{equation}
with $\delta$ being the Doppler factor, whereas an alternative definition of $\delta$ and the $\Gamma$ the bulk Lorentz factor are given by:
\begin{equation}
    \delta = \frac{1}{\Gamma (1 - \beta \cos \theta) }  , ~ ~ \Gamma = (1-\beta^2)^{-1/2}.
    \label{eq:delta}
\end{equation}

\noindent By adopting a mean variability Doppler factor for the entire jet of $\delta = 32$ \citep{2017ApJ...846...98J} we derived $T_\mathrm{b,int}$ from $\sim 0.003 \times 10^{10}$\,K to $\sim 28\times 10^{10}$\,K at 43\,GHz. The wide range observed suggests variations in the energy distribution and properties of the emitting plasma along the 43\,GHz jet. Specifically, brightness temperature values below the equipartition limit ($T_\mathrm{b,eq}$) suggest regions where the magnetic field strength dominates over the particle energy. In such regions, the synchrotron radiation from the electrons is suppressed due to the presence of a strong magnetic field. Conversely, brightness temperature values exceeding the $T_\mathrm{b,eq}$ limit indicate particle-dominated regions. At 86\,GHz, the intrinsic brightness temperatures range from approximately $\sim 0.006 \times 10^{10}$\,K to $\sim 5\times 10^{10}$\,K, indicating a magnetically dominated jet.

\begin{table}
\caption{Kinematic parameters of the identified jet components at 43\,GHz, and 86\,GHz. } 
\label{table:speeds}
\centering
\begin{tabular}{@{}lccr@{\,$\pm$\,}lr@{\,$\pm$\,}lr@{\,$\pm$\,}l@{}}
\hline\hline   
Knot                & Freq. & N  & \multicolumn{2}{c}{$\mu$}           & \multicolumn{2}{c}{$\beta_\mathrm{app}$}          & \multicolumn{2}{c}{$\theta_\mathrm{c}$}        \\
                    & (GHz) &    & \multicolumn{2}{c}{(mas/year)}      & \multicolumn{2}{c}{(c)}                    & \multicolumn{2}{c}{($^{\circ}$)}             
\\
(1) & (2) & (3) & \multicolumn{2}{c}{(4) } & \multicolumn{2}{c}{(5)} & \multicolumn{2}{c}{(6)} \\ \hline
\multirow{2}{*}{J9} & 43    & 4 & 0.3&0.1 & 15&5      & 3.9&1.4 \\
                    & 86    & 2  & 0.2&0.2 & 10&9     & 5.8&5.8          \\
                    \hline
\multirow{2}{*}{J8} & 43    & 7  & 0.5&0.1 & 21&5  & 2.7&1.2  \\
                    & 86    & 3  & 0.3&0.1 & 15&2  & 3.9&0.6  \\
                    \hline
\multirow{2}{*}{J6} & 43    & 2  & 0.6&1.0 & 26&42    & 2.2&3.5           \\
                    & 86    & 2  & 0.4&0.1 & 20&3    & 2.9&0.4   \\
 \hline
\multirow{2}{*}{J5} & 43    & 3  & 0.4&0.1 & 19&1     & 3.0&0.1  \\
                    & 86    & 2  & 0.6&0.2 & 30&8     & 2.0&0.1 \\
\hline
\multirow{2}{*}{J4} & 43    & 4  & 0.5&0.1  & 25&5     & 2.4&0.5  \\
                    & 86    & 2  & 0.5&0.1 & 24&2     & 2.4&0.2  \\
                    \hline
\multirow{2}{*}{M} & 43    & 8   & 0.5&0.1 & 22&1     & 2.7&0.1  \\
                    & 86   & 3   & 0.4&0.1 & 19&1      & 3.0&0.2  \\
                    \hline
J1                  & 43   & 4   & 0.3&0.1 & 12&4      & 4.5&1.3   \\
\hline  
\end{tabular}
\tablefoot{Columns from left to right: (1) component ID, (2) observing frequency, (3) number of data points, (4) proper motion, (5) apparent speed, and (6) viewing angle.}
\end{table}

\section{Discussion}
\label{sec:discussion}

Jets emitted by AGN have been extensively studied with VLBI observations, revealing the presence of "stationary" or "quasi-stationary" features in the jet \citep[e.g.,][and references therein]{2022ApJS..260...12W}. The exact nature of these features has been the subject of various explanations, including shock waves, standing recollimation shocks, sites of maximised Doppler beaming, and stationary shocks. These scenarios aim to explain how the jet material is suddenly decelerated and compressed, or where the jet abruptly bends.
The jet in 3C\,345.3 exhibits a unique and prominent feature, region C, which is stationary and appears to be the terminal point for all moving knots after the middle of 2014. Our goal is to explore the aforementioned scenarios in order to provide a comprehensive explanation for this phenomenon.

\subsection{Region C as a recollimation or standing shock}

One hypothesis that may explain the peculiar behavior observed in region C is the presence of a recollimation shock. One of the most prevalent factors influencing the dynamics and stability of extragalactic jets is the mismatch in pressure between the jet and its surrounding medium. Such pressure mismatches result in the expansion and recollimation of the jet, forming conical shocks, and, under high-pressure conditions, Mach disks and decollimation shocks \citep[e.g.,][]{2013EPJWC..6102002P}. When examining VLBI images, the manifestation of such a shock is expected to resemble a quasi-stationary feature, similar to the observations made in region C.

Numerical simulations of the relativistic hydrodynamics and emission of jets \citep{1995ApJ...449L..19G,1997ApJ...482L..33G} show that as moving ejecta traverse such a shock, the two components appear to merge into a single one, causing the position of the merged feature to shift downstream relative to the original location of the stationary knot. 
After the collision, the two components separate, with the quasi-stationary feature reverting to its initial position while the moving feature continues its trajectory. \cite{Gomez_1999} reported such behavior in region C.
\\
\indent Another observational signature of a recollimation shock can be the increase in the brightness temperature. Bright stationary components associated with the recollimation shocks appear presenting a relative intensity modulated by the Doppler boosting ratio between the pre-shock and post-shock states. This can lead to an increase in brightness temperature \citep{2008ASPC..386..488R,Fuentes_2018}. Also, in the case of strong shock-shock interactions, the formation of relaxation shocks behind the leading event, can lead to an increase in brightness temperature \citep{2018A&A...610A..32B,2022A&A...661A..54F}.
\\
\indent Recollimation shocks can induce alterations in the polarization of the radio emission from the jet too \citep{2013ApJ...772...14C, Cawthorne}. In particular, the shock can compress and reorient the magnetic field within the jet, resulting in modifications to both the orientation and degree of polarization of the emitted radiation. The presence of an ordered magnetic field perpendicular to the bulk jet flow is the most common observational signature of a recollimation shock. 
\\
\indent The polarized images of \object{3C\,454.3} at 43\,GHz and 86\,GHz reveal EVPAs that are aligned with the jet axis, assuming a linear plasma flow through region C. This alignment suggests a predominantly toroidal magnetic field in the jet, supporting the recollimation shock scenario. However, even in the presence of a Mach disk, which has the potential to slow down or momentarily disrupt the jet flow \citep{2007MNRAS.382..526P}, a recollimation shock is unable to halt a relativistically moving disturbance along the jet, as observed in \object{3C\,454.3}. The same applies also for the case that region C corresponds to a standing shock. Therefore, we are limited in alternative physical interpretations or, alternatively, we ought to consider a superposition of physical phenomena.

\subsection{Region C as a jet bend}

Another potential explanation for the quasi-stationarity of knots observed in region C may involve a segment of the jet that aligns with our line of sight. Such a geometrical coincidence could induce an increase in flux density due to Doppler boosting, coupled with an EVPA rotation, as evidenced in the \cite{1994A&A...284...51G,1994A&A...292...33G}, and similar with what we see in knot M (\autoref{fig:vlbi_maps}, and column 4 in \autoref{table:kinem43} and \autoref{table:kinem86}). In this case, we would anticipate a rise in the T$_\mathrm{b,obs}$ by a factor of $(\delta_\mathrm{new}/\delta_\mathrm{old})^{n}$, where n is a dimensionless exponent associated with the spectral index of the jet ($2\!-\!\alpha$ for the case of a continuous jet flow or $3\!-\!\alpha$ for a moving inhomogeneity; \citet{1994ApJ...426...51R}). Indeed, after the arrival of M, the $T_\mathrm{b,obs}$ of knots in region C show an extreme deviation from the expected T$_\mathrm{b,obs}$ (red dotted line in \autoref{fig:tb}). For estimating the required $\delta_\mathrm{new}$, we go on to consider knot M as a moving plasma inhomogeneity, we adopt a typical spectral index of $\alpha=-1$ (which results in n=4), and by setting $\delta_\mathrm{old}=21.5 \pm 0.5$ (based on the apparent speed of the most well-defined component in our sample, the Doppler factor of M at 43\,GHz computed by $\delta_\mathrm{app,M} \approx \gamma_\mathrm{min,M}=\sqrt{\beta_\mathrm{app,M}^{2}+1}$), we obtain in the location of the maximum $T_\mathrm{b,obs}$ for Ca: $\delta_\mathrm{new,Ca}= 60.4 \pm 4.5$, for Cb: $\delta_\mathrm{new,Cb}= 103.7 \pm 6.6$, and for Cc: $\delta_\mathrm{new,Cc}= 81.1 \pm 5.6$. Next, using these values and the minimum Lorentz factor of M as the bulk jet speed ($\gamma_\mathrm{min,M}= \Gamma=22$, assuming that it does not change dramatically along the bend), we use \autoref{eq:delta} to show that for a complete alignment with our line of sight, the maximum  $\delta_\mathrm{new}$ is $\approx$44 for all Ca, Cb, and Cc. Therefore, the increase we measure of $T_\mathrm{b,obs}$ cannot be caused only by a jet bending, but a combination with another process that increases the intrinsic energetic of the jet is required, like jet flow acceleration, unusually large particle acceleration, and/or increment of particle density. 

Additionally, if bending towards the line of sight were the primary mechanism explaining the slow-down of component motion, one would anticipate a shift towards a less steep spectral index due to increased path lengths and higher opacity. Direct observation into the jet funnel would result in a not-so-steep spectral index. However, the observation of a very steep spectral index suggests optically thin emission. This is further corroborated by the consistent and high polarization in region C and similar EVPAs at 43 and 86 GHz, indicative of a low Faraday depth.

Nonetheless, as a strong jet bending seems most plausible (supported also by the spectral index findings in \Cref{sec:spix}), we propose a geometrical model where at the location of component Ca the jet takes its first gentle turn towards our line of sight, achieves complete alignment of the plasma flow at component Cb, and then diverts away from us at component Cc. Alternatively, the jet might exhibit a single substantial bend, with a small section parallel to our line of sight, appearing the brightest due to the acute viewing angle of the remainder of the bend. The aforementioned scenarios imply that the path of M changed at least from $\theta_\mathrm{c}=2.7^\circ$ to $0^\circ$. 

\cite{2005AJ....130.1418J}, based on VLBI polarimetric observations at 43\,GHz, presented evidence of feature C lying very close to our line of sight, and that moving knots which are detected beyond C, appear to follow trajectories that correspond to two different groups: a northwestern and a southwestern one. The projected difference between those trajectories is measured to be 70$^{\circ}$. This particular result is consistent with the scenario in which the direction of the jet near C is lying very close to our line of sight. Such geometry can introduce substantial differences in projected trajectories of moving components, even if the intrinsic trajectories differ only slightly.

If this hypothesis is valid, a systematic change in the apparent motion of all components that are passing through this region would be expected to be seen. Indeed, beyond region C the components become invisible. Additionally, a jet bend would also significantly increase otherwise minor projection effects, which means that while component M is moving along the bend, it appears to be stretched out, therefore being resolved into three different sub-components, namely Ca, Cb, and Cc. Nevertheless, the alternative scenarios of a single very large bend can be also considered, with a small section of this bend to be parallel to our line of sight, whereas the rest of the bend would be observed at very large viewing angle. 

Observationally, similar jet features, such as region C, are frequently associated with spatially bent (and possibly helical) jet structures in which the jet Lorentz factor remains constant along the outflow but the jet viewing angle varies \citep[e.g.,][]{2010ApJ...712L.160R,2014A&A...567A..41A}. Jet bends on VLBI scales are a common feature in many blazar jets \citep[e.g.,][]{1989A&A...211L..23M,1994cers.conf...39K,2005A&A...431..831L,2011A&A...529A.113Z,2012ApJ...749...55P}. For relativistic jets, the small angle of the jet axis to the line-of-sight amplifies the bend angle. A similar geometric effect was found in the blazar 4C\,39.25, where two nearby stationary hotspots in the jet, "a" and "c," correspond to two jet bends that redirect the plasma flow towards our line of sight, with the latter returning to its initial direction \citep{1993ApJ...402..160A}. However, between features a and c the spectral evolution of a transverse shock, which was propagating down the curving jet, helped to decipher the nature of the standing features; something we cannot detect in our case. Another example is the blazar 3C\,279, which exhibits an elongated nuclear structure perpendicular to the large-scale jet \citep{2020A&A...640A..69K}. Three bright features separated by $0.3-0.4$,mas were used to model the structure, which is comparable to Ca, Cb, and Cc in our case. The authors suggest that this phenomenology can be explained by a jet closely aligned to the line of sight, with propagating knots moving along the bend with the same bulk Lorentz factors but different viewing angles. Finally, in order to explain the sparsity of the measured apparent speed values, a strong jet bending away from the line of sight is reported also in the blazar J1924-2914 \citep{2022ApJ...934..145I}.

\subsubsection{Physical origin of the bending}

Various physical mechanisms can contribute in such an extreme bending, one of which is large-scale magnetic reconnection. Indeed, this magnetic configuration can give rise to the formation of elongated structures, such as linear strings of "plasmoids" \citep{2017SSRv..207..291B}, or oblique structures like shocks or jet filaments \citep{1985ApJ...298..114M}. These structures can be easily produced by the extension of magnetic field loops due to cross-jet velocity gradients. In this scenario, we would expect the formation of elongated loops with field lines pointing in opposite directions. Interestingly, in our case, the projected EVPAs are indeed perpendicular to the elongated emission structure, considering that the plasma flow after the location of the component Ca moves from north to south. Alternatively, in the case of a supersonic flow, an oblique standing shock would also decrease the component of the flow velocity parallel to the shock normal causing the flow to bend. This would lead to compression of the magnetic field and particle acceleration, in addition to a change in the Doppler factor \citep{1990ApJ...350..536C,1997A&A...327..550T}. 
\\
\indent Local density enhancements can also be triggered by plasma instabilities during the jet propagation \citep[e.g.,][]{1984ApJ...287..523H,1984ApJ...277..106H}. In the linear perturbation regime, instabilities are likely to occur in relativistic plasma flows, although it remains uncertain whether their amplitudes can grow sufficiently to induce a significant bend in the entire jet. The non-axisymmetric helical mode of the Kelvin-Helmholtz (KH) instability, which can arise from a velocity shear between two plasma layers within the jet \citep[][and references therein]{Mizuno_2007}, has the potential to create bent structures in relativistic jets \citep{2000ApJ...533..176H}. Alternatively, the current-driven (CD) kink instability has been shown to produce helically twisted jet structures \citep{2009MNRAS.394L.126M,Mizuno_2012}. Although it is not necessarily true that the jet axis itself bends, plasma disturbances may move along different filaments inside the funnel. \cite{1990ApJ...365..134H} showed that macroscopic fluid instabilities can cause a jet to bend. Numerical simulations have shown \citep{2006A&A...456..493P} that an asymmetric perturbation at the base of a relativistic jet propagates downstream and naturally creates a pressure maximum helical structure, which can eventually distort the jet flow and force it into a helical path when its amplitude grows to large enough values. 
\\
\indent A bend in the jet could also be attributed to changes in the orientation of the central engine and the jet nozzle. While individual elements of the jet may be moving in linear trajectories, the apparent bending arises from the successive elements changing their directions. A specific case related to this scenario involves precession, where the jet undergoes a rotational motion. Fluctuations in the magnetic field in the inner disk, variations in the accretion flow, or the development of shocks or instabilities in the jet can induce precession. Recent studies have suggested the potential existence of a very massive binary black hole system, with a possible period of 14 years, making it a strong emitter of gravitational waves \citep{2021A&A...648A..27V, 2021A&A...653A...7Q}. Within this framework, the observed variability in \object{3C\,454.3} can be attributed to the consequences of jet precession.
\\
\indent Jet-cloud collisions can induce structural changes in the jet, potentially causing disruption or deflection due to cloud interactions. A study of the jet head position flip in 3C\,84 in 2015 provided evidence for a strong jet-cloud collision that triggered the observed change. \cite{Kino_2021} suggested that the collision resulted in a shock wave propagating through the cloud, leading to magnetic field compression and increased magnetic pressure. This, in turn, caused the jet to become collimated and accelerated, leading to the observed morphological transition of the compact radio lobe in 3C\,84. 
\\
\indent Similar collisions have also been reported in the Seyfert galaxy NGC\,3079 \citep{2007MNRAS.377..731M}, the quasar 3C\,279, and the radio galaxy 3C\,120 \citep[][and referances therein]{2008ASPC..386..240L,2011ApJ...733...11G}. \cite{2010A&A...522A..97A} investigating the interaction of clouds with the base of AGN jets concluded that clouds can only enter the jet at a certain height. Below this height, the jet is too compact and its ram and magnetic pressure will destroy the cloud before it fully penetrates the jet. An application of this model to the blazar 3C\,273 showed that jet-cloud interactions can indeed occur in this source. A similar case can also trigger the observed behavior in region C. The jet of \object{3C\,454.3} could potentially be stopped or deviated by a large cloud, leading to the jet upstream of region C to be less Doppler-boosted and, therefore, invisible to us. Even though little is known about the external medium surrounding \object{3C\,454.3}, partly due to the lack of observations of extended X-ray emission near the source, we can estimate the height at which clouds can enter the jet and compare it with the location of region C. 
\\
\indent With a jet luminosity of $L_\mathrm{j} \approx 10^{47}$\,erg s$^{-1}$ \citep{2011MNRAS.410..368B} and by adopting a typical lifetime and speed of the clouds, the minimum height above which clouds can enter the jet is $\approx$0.10\,pc. In contrast, region C is located much further downstream at a projected distance of $\approx$4.5\,pc, supporting the possibility of jet-cloud interactions.

\section{Conclusions}

In this work, we present high-resolution VLBI polarimetric images of blazar \object{3C\,454.3}. The images were obtained with the GMVA at 86\,GHz from September 2013 until March 2017.
Combining 43\,GHz VLBA observations from the VLBA-BU BLAZAR program with our 86\,GHz GMVA data, we study the kinematics of the source, revealing the disappearance of moving jet components entering region C. Several scenarios have been considered for explaining this behaviour, including the presence of a recollimation or standing shock and a jet bend. Overall, the observations and analysis presented in this study suggest that the peculiar behavior of region C can be explained by a combination of a jet bending and a moving shock and, perhaps, some jet flow acceleration and/or unusually large particle acceleration as well. Further observations are needed to fully understand the physical processes and dynamics involved in the formation of jet bends and their interactions with the surrounding environment.

\begin{acknowledgements}

I would like to express my gratitude to the Professors Pantelis Papadopoulos and Nikolaos Stergioulas from the Aristotle University of Thessaloniki (APTH), as well as the post-doctoral researcher Nikos Karnesis for the fruitful discussions and their hospitality.

This research has made use of data obtained with the Global Millimeter VLBI Array (GMVA), which consists of telescopes operated by the MPIfR, IRAM, Onsala, Mets\"ahovi, Yebes, the Korean VLBI Network, the Green Bank Observatory and the Very Long Baseline Array (VLBA). The VLBA is a facility of the National Science Foundation operated under cooperative agreement by Associated Universities, Inc. The data were correlated at the DiFX correlator of the MPIfR in Bonn, Germany. We express our special thanks to the people supporting the observations at the telescopes during the data collection.We express our special thanks to the observers at the telescopes and the observatory staff for their help and support in running the GMVA.

This research has made use of data from the OVRO 40-m monitoring program (Richards, J. L. et al. 2011, ApJS, 194, 29), which is supported in part by NASA grants NNX08AW31G, NNX11A043G, and NNX14AQ89G and NSF grants AST-0808050 and AST-1109911.

The Submillimeter Array is a joint project between the Smithsonian Astrophysical Observatory and the Academia Sinica Institute of Astronomy and Astrophysics and is funded by the Smithsonian Institution and the Academia Sinica.

This research has made use of NASA's Astrophysics Data System.

Author Efthalia Traianou acknowledges financial support from the grant CEX2021-001131-S funded by MCIN/AEI/ 10.13039/501100011033.

This work is partly based on observations carried out with the IRAM 30m telescope. IRAM is supported by INSU/CNRS (France), MPG (Germany) and IGN (Spain).

JYK was supported for this research by the National Research Foundation of Korea (NRF) funded by the Korean government (Ministry of Science and ICT; grant no. 2022R1C1C1005255).

C.C. acknowledges support by the European Research Council (ERC) under the HORIZON ERC Grants 2021 programme under grant agreement No. 101040021.

This work is part of the M2FINDERS project which has received funding from the European Research Council (ERC) under the European Union’s Horizon 2020 Research and Innovation Programme (grant agreement No 101018682).

The research at Boston University was supported in part by NASA Fermi GI grant 80NSSC20K1567.

\end{acknowledgements}

\bibliographystyle{aa} 
\bibliography{aanda}

\begin{appendix}

\section{Model-fitting parameters}

\begin{table*}
\caption{Model-fitting parameters at 43\,GHz.}
\label{table:kinem43}
\centering
\begin{tabular}{cccccccc}
\hline\hline  
Knot                   & Region C             & Time                           & S    & RA    & DEC   & FWHM   & $T_b$      \\
                       &                      & (Years)                        & (Jy) & (mas) & (mas) & (mas) & (10$^{\rm{10}}$K) \\
(1) & (2) & (3) & (4) & (5) & (6) & (7) & (8) \\ \hline
\multirow{24}{*}{Core}  &                          & 2013.57                        & 5.13    $\pm$ 0.50  & $-$            & $-$            & 0.05   $\pm$    0.002  & 251.7       $\pm$       24.53      \\
                        &                          & 2013.65                        & 5.59    $\pm$ 0.60  & $-$           & $-$            & 0.05   $\pm$    0.002  & 274.27      $\pm$       29.44      \\
                        &                          & 2013.88                        & 8.90    $\pm$ 0.90  & $-$            & $-$            & 0.05   $\pm$    0.003  & 436.67      $\pm$       44.16      \\
                        &                          & 2014.15                        & 11.01   $\pm$ 1.10  & $-$            & $-$            & 0.06   $\pm$    0.002  & 375.13      $\pm$       37.48      \\
                        &                          & 2014.33 & 16.10   $\pm$ 1.60  & $-$            & $-$            & 0.09   $\pm$    0.003  & 243.81      $\pm$       24.23      \\
                        &                          & 2014.47 & 23.60   $\pm$ 2.40  & $-$            & $-$            & 0.12   $\pm$    0.004  & 201.03      $\pm$       20.44      \\
                        &                          & 2014.57 & 5.74    $\pm$ 0.60  & $-$            & $-$            & 0.10    $\pm$    0.006  & 70.41       $\pm$       7.36       \\
                        &                          & 2014.72                        & 6.29    $\pm$ 0.60  & $-$            & $-$           & 0.10    $\pm$    0.004  & 77.15       $\pm$       7.36       \\
                        &                          & 2014.99                        & 3.86    $\pm$ 0.40  & $-$           & $-$           & 0.08   $\pm$    0.005  & 73.98       $\pm$       7.67       \\
                        &                          & 2015.12                        & 3.50    $\pm$ 0.35  & $-$           & $-$            & 0.10    $\pm$    0.003  & 42.93       $\pm$       4.29       \\
                        &                          & 2015.28                        & 4.79    $\pm$ 0.50  & $-$          & $-$           & 0.14   $\pm$    0.01   & 29.98       $\pm$       3.13       \\
                        &                          & 2015.36                        & 2.11    $\pm$ 0.21  & $-$           & $-$           & 0.11   $\pm$    0.01   & 21.39       $\pm$       2.13       \\
                        &                          & 2015.44                        & 1.84    $\pm$ 0.20  & $-$           & $-$            & 0.06   $\pm$    0.005  & 46.06       $\pm$       5.01       \\
                        &                          & 2015.58                        & 2.25    $\pm$ 0.25  & $-$           & $-$            & 0.06   $\pm$    0.004  & 76.66       $\pm$       8.52       \\
                        &                          & 2015.99                        & 1.94    $\pm$ 0.20  & $-$          & $-$            & 0.07   $\pm$    0.002  & 48.56       $\pm$       5.01       \\
                        &                          & 2016.08                        & 2.71    $\pm$ 0.30  & $-$          & $-$            & 0.07   $\pm$    0.002  & 67.84       $\pm$       7.51       \\
                        &                          & 2016.31                        & 3.16    $\pm$ 0.31  & $-$            & $-$            & 0.06   $\pm$    0.002  & 107.67      $\pm$       10.56      \\
                        &                          & 2016.44                        & 2.10     $\pm$ 0.21  & $-$           & $-$            & 0.05   $\pm$    0.004  & 103.03      $\pm$       10.3       \\
                        &                          & 2016.68                        & 4.46    $\pm$ 0.45  & $-$          & $-$            & 0.06   $\pm$    0.005  & 151.96      $\pm$       15.33      \\
                        &                          & 2016.81                         & 4.60     $\pm$ 0.46  & $-$           & $-$            & 0.07   $\pm$    0.005  & 115.15      $\pm$       11.51      \\
                        &                          & 2016.90                         & 3.74    $\pm$ 0.38  & $-$          & $-$             & 0.04   $\pm$    0.002  & 286.72      $\pm$       29.13      \\
                        &                          & 2016.97                        & 6.23    $\pm$ 0.62  & $-$          & $-$           & 0.06   $\pm$    0.001  & 212.27      $\pm$       21.12      \\
                        &                          & 2017.09                        & 10.70   $\pm$ 1.10  & $-$          & $-$            & 0.04   $\pm$    0.02   & 820.29      $\pm$       84.33      \\
     &                          & 2017.29                        & 8.16    $\pm$ 0.82  & $-$       & $-$           & 0.08   $\pm$    0.02   & 156.39      $\pm$       15.72      \\
     \hline
    & \multirow{11}{*}{C}                       & 2013.57                        & 0.58    $\pm$ 0.06  & $-$0.40  $\pm$ 0.05    & $-$0.01   $\pm$ 0.05   & 0.26   $\pm$    0.07   & 1.05        $\pm$       0.11       \\
                        &                          & 2013.65                        & 0.72    $\pm$ 0.07  & $-$0.44  $\pm$ 0.08    & $-$0.03   $\pm$ 0.08   & 0.40   $\pm$    0.10   & 0.55        $\pm$       0.06       \\
                        &                          & 2013.88                        & 0.58    $\pm$ 0.06  & $-$0.41  $\pm$ 0.06    & $-$0.01   $\pm$ 0.06   & 0.32   $\pm$    0.12   & 0.69        $\pm$       0.07       \\
                        &                          & 2014.15                        & 0.68    $\pm$ 0.07  & $-$0.45  $\pm$ 0.09    & $-$0.05   $\pm$ 0.09   & 0.47   $\pm$    0.002  & 0.38        $\pm$       0.04       \\
                        &                          & 2014.34                        & 1.21    $\pm$ 0.12  & $-$0.29  $\pm$ 0.05    & $-$0.06   $\pm$ 0.05   & 0.24   $\pm$    0.06   & 2.58        $\pm$       0.26       \\
                        &                          & 2014.47 & 1.46    $\pm$ 0.15  & $-$0.32  $\pm$ 0.03    & $-$0.07   $\pm$ 0.03   & 0.28   $\pm$    0.07   & 2.28        $\pm$       0.24       \\
                        &                          & 2014.57                        & 0.35    $\pm$ 0.04  & $-$0.47  $\pm$ 0.09    & $-$0.13   $\pm$ 0.09   & 0.44   $\pm$    0.25   & 0.20         $\pm$       0.03       \\
                        &                          & 2014.72                        & 0.35    $\pm$ 0.04  & $-$0.49  $\pm$ 0.05    & $-$0.09   $\pm$ 0.05   & 0.29   $\pm$    0.11   & 0.45        $\pm$       0.05       \\
                        &                          & 2014.99                        & 0.17    $\pm$ 0.02  & $-$0.33  $\pm$ 0.05    & $-$0.10   $\pm$ 0.05   & 0.26   $\pm$    0.06   & 0.31        $\pm$       0.04       \\
                        &                          & 2015.12                        & 0.33    $\pm$ 0.33  & $-$0.58  $\pm$ 0.10     & $-$0.37   $\pm$ 0.10    & 0.45   $\pm$    0.11   & 0.20         $\pm$       0.20        \\
   &                          & 2015.99                        & 0.26    $\pm$ 0.03  & $-$0.69  $\pm$ 0.05    & 0.01    $\pm$ 0.05   & 0.23   $\pm$    0.09   & 0.25        $\pm$       0.03       \\
\hline
                        & \multirow{11}{*}{C$_a$}                     & 2015.58                        & 3.29    $\pm$ 0.33  & $-$0.49  $\pm$ 0.02    & 0.23    $\pm$ 0.02   & 0.12   $\pm$    0.01  & 28.02       $\pm$       2.81       \\
                        &                          & 2015.99                        & 2.62    $\pm$ 0.26  & $-$0.52  $\pm$ 0.03    & 0.26    $\pm$ 0.03   & 0.15   $\pm$    0.01  & 14.28       $\pm$       1.42       \\
                        &                          & 2016.08                        & 3.15    $\pm$ 0.31  & $-$0.52  $\pm$ 0.03    & 0.26    $\pm$ 0.03   & 0.17   $\pm$    0.02   & 13.37       $\pm$       1.32       \\
                        &                          & 2016.31                        & 2.77    $\pm$ 0.30  & $-$0.5   $\pm$ 0.04    & 0.26    $\pm$ 0.04   & 0.22   $\pm$   0.02   & 7.02        $\pm$       0.76       \\
                        &                          & 2016.44                        & 2.90     $\pm$ 0.30  & $-$0.45  $\pm$ 0.07    & 0.22    $\pm$ 0.07   & 0.33   $\pm$    0.03   & 3.27        $\pm$       0.34       \\
                        &                          & 2016.68                        & 3.21    $\pm$ 0.32  & $-$0.52  $\pm$ 0.04    & 0.19    $\pm$ 0.04   & 0.20    $\pm$    0.02   & 9.84        $\pm$       0.98       \\
                        &                          & 2016.80                         & 1.01    $\pm$ 0.10  & $-$0.44  $\pm$ 0.04    & 0.29    $\pm$ 0.04   & 0.21   $\pm$    0.01   & 2.81        $\pm$       0.28       \\
                        &                          & 2016.90                         & 0.89    $\pm$ 0.90  & $-$0.44  $\pm$ 0.04    & 0.22    $\pm$ 0.04   & 0.18   $\pm$    0.02   & 3.37        $\pm$       3.41       \\
                        &                          & 2016.97                        & 1.51    $\pm$ 0.15  & $-$0.56  $\pm$ 0.04    & 0.24    $\pm$ 0.04   & 0.22   $\pm$    0.02   & 3.83        $\pm$       0.38       \\
\hline
                        & \multirow{11}{*}{C$_b$}    & 2015.58                        & 8.01    $\pm$ 0.80  & $-$0.54  $\pm$ 0.02    & 0.20     $\pm$ 0.02   & 0.10   $\pm$    0.002  & 98.25       $\pm$       9.81       \\
                        &                          & 2015.99                        & 5.09    $\pm$ 0.51  & $-$0.60   $\pm$ 0.02    & 0.16    $\pm$ 0.02   & 0.06   $\pm$    0.003  & 173.43      $\pm$       17.38      \\
                        &                          & 2016.08                        & 4.45    $\pm$ 0.45  & $-$0.57  $\pm$ 0.01    & 0.15    $\pm$ 0.01   & 0.04   $\pm$    0.004  & 341.15      $\pm$       34.5       \\
                        &                          & 2016.31                        & 4.24    $\pm$ 0.42  & $-$0.59  $\pm$ 0.01    & 0.11    $\pm$ 0.01   & 0.06   $\pm$    0.001  & 144.47      $\pm$       14.31      \\
                        &                          & 2016.44                        & 3.37    $\pm$ 0.34  & $-$0.57  $\pm$ 0.02    & 0.16    $\pm$ 0.02   & 0.08   $\pm$    0.005  & 64.59       $\pm$       6.52       \\
                        &                          & 2016.68                        & 4.92    $\pm$ 0.50  & $-$0.59  $\pm$ 0.02    & 0.06    $\pm$ 0.02   & 0.08   $\pm$    0.005  & 94.29       $\pm$       9.58       \\
                        &                          & 2016.80                        & 3.17    $\pm$ 0.32  & $-$0.58  $\pm$ 0.02    & 0.15    $\pm$ 0.02   & 0.11   $\pm$    0.006  & 32.13       $\pm$       3.24       \\
                        &                          & 2016.90                        & 2.61    $\pm$ 0.62  & $-$0.58  $\pm$ 0.02    & 0.14    $\pm$ 0.02   & 0.10   $\pm$    0.010  & 32.01       $\pm$       7.6        \\
                        &                          & 2016.97                        & 3.80    $\pm$ 0.38  & $-$0.57  $\pm$ 0.02    & 0.08    $\pm$ 0.02   & 0.10   $\pm$    0.010  & 46.61       $\pm$       4.66       \\
                        &                          & 2017.09                        & 4.80    $\pm$ 0.48  & $-$0.56  $\pm$ 0.03    & 0.11    $\pm$ 0.03   & 0.16   $\pm$    0.017  & 23.0        $\pm$       2.3        \\
                        &  & 2017.29                        & 1.76    $\pm$ 0.18  & $-$0.52  $\pm$ 0.03    & 0.18    $\pm$ 0.03   & 0.16   $\pm$    0.017  & 8.43        $\pm$       0.86       \\
\hline  
\end{tabular}
\tablefoot{Columns from left to right: (1) component ID, (2) component ID in region C, (3) observed epoch, (4) flux density, (5) relative right ascension, (6) relative declination, (7) component size, and (8) brightness temperature.}
\label{table:knots43a}
\end{table*}

\addtocounter{table}{-1} 
\begin{table*}
\caption{Model-fitting parameters at 43\,GHz continued.}
\centering
\begin{tabular}{cccccccc}
\hline\hline  
Knot                   & Region C             & Time                           & S    & RA    & DEC   & FWHM   & $T_b$      \\
                       &                      & (Years)                        & (Jy) & (mas) & (mas) & (mas) & (10$^{\rm{10}}$K) \\
(1) & (2) & (3) & (4) & (5) & (6) & (7) & (8) \\ \hline
                        &  \multirow{11}{*}{C$_c$}      & 2015.58                        & 2.15    $\pm$ 0.21  & $-$0.52  $\pm$ 0.03    & 0.04    $\pm$ 0.03   & 0.17   $\pm$    0.01   & 9.13        $\pm$       0.89       \\
                        &                          & 2015.99                        & 2.70    $\pm$ 0.27  & $-$0.54  $\pm$ 0.04    & 0.02    $\pm$ 0.04   & 0.21   $\pm$    0.01   & 7.51        $\pm$       0.75       \\
                        &                          & 2016.08                        & 1.90    $\pm$ 0.20  & $-$0.52  $\pm$ 0.04    & $-$0.03   $\pm$ 0.04   & 0.18   $\pm$    0.02   & 7.19        $\pm$       0.76       \\
                        &                          & 2016.31                        & 2.89    $\pm$ 0.30  & $-$0.56  $\pm$ 0.03    & $-$0.01   $\pm$ 0.03   & 0.15   $\pm$    0.01   & 15.75       $\pm$       1.64       \\
                        &                          & 2016.44                        & 3.69    $\pm$ 0.37  & $-$0.59  $\pm$ 0.02    & 0.02    $\pm$ 0.02   & 0.08   $\pm$    0.01   & 70.72       $\pm$       7.09       \\
                        &                          & 2016.68                        & 2.16    $\pm$ 0.22  & $-$0.55  $\pm$ 0.02    & $-$0.07   $\pm$ 0.02   & 0.11   $\pm$    0.01   & 21.9        $\pm$       2.23       \\
                        &                          & 2016.80                        & 2.59    $\pm$ 0.26  & $-$0.57  $\pm$ 0.03    & $-$0.03   $\pm$ 0.03   & 0.15   $\pm$    0.01   & 14.12       $\pm$       1.42       \\
                        &                          & 2016.90                        & 2.12    $\pm$ 0.21  & $-$0.56  $\pm$ 0.03    & $-$0.04   $\pm$ 0.03   & 0.14   $\pm$    0.01   & 13.27       $\pm$       1.31       \\
                        &                          & 2016.97                        & 1.40    $\pm$ 0.14  & $-$0.57  $\pm$ 0.04    & $-$0.12   $\pm$ 0.04   & 0.19   $\pm$    0.01   & 4.76        $\pm$       0.48       \\
                        &                          & 2017.09                        & 1.13    $\pm$ 0.11  & $-$0.56  $\pm$ 0.03    & $-$0.11   $\pm$ 0.03   & 0.15   $\pm$    0.02   & 6.16        $\pm$       0.6        \\
                        &   & 2017.29                        & 1.77    $\pm$ 0.20  & $-$0.49  $\pm$ 0.04    & $-$0.02   $\pm$ 0.04   & 0.19   $\pm$    0.02   & 6.01        $\pm$       0.68       \\
\hline
\multirow{4}{*}{J1}                        &                          & 2013.57                        & 0.49    $\pm$ 0.05  & $-$0.68  $\pm$ 0.14  & $-$0.15   $\pm$ 0.14   & 0.49   $\pm$    0.14   & 0.25        $\pm$       0.03       \\
                        &                          & 2013.65                        & 0.14    $\pm$ 0.01  & $-$0.76  $\pm$ 0.05  & $-$0.23   $\pm$ 0.05   & 0.27   $\pm$    0.10   & 0.16        $\pm$       0.01       \\
                        &                          & 2013.88                        & 0.37    $\pm$ 0.04  & $-$0.73  $\pm$ 0.15  & $-$0.21   $\pm$ 0.15   & 0.46   $\pm$    0.15   & 0.21        $\pm$       0.03       \\
    &                          & 2014.15                        & 0.12    $\pm$ 0.01  & $-$0.86  $\pm$ 0.07    & $-$0.20    $\pm$ 0.07   & 0.35   $\pm$    0.12   & 0.12        $\pm$      0.01       \\
\hline
\multirow{8}{*}{M}                        &                          & 2014.47 & 6.53    $\pm$ 0.65  & $-$0.05  $\pm$ 0.004   & 0.07    $\pm$ 0.004  & 0.02   $\pm$    0.005  & 889.97      $\pm$       88.59      \\
                        &                          & 2014.57                        & 7.40     $\pm$ 0.74  & $-$0.11  $\pm$ 0.01    & 0.05    $\pm$ 0.01   & 0.07   $\pm$    0.01   & 185.24      $\pm$       18.52      \\
                        &                          & 2014.72                        & 8.80    $\pm$ 0.88  & $-$0.15  $\pm$ 0.02    & 0.08    $\pm$ 0.02   & 0.07   $\pm$   0.01   & 220.29      $\pm$       22.03      \\
                        &                          & 2014.99                        & 4.50    $\pm$ 0.45  & 0.16   $\pm$ 0.02    & $-$0.05   $\pm$ 0.02   & 0.10    $\pm$    0.02   & 55.2        $\pm$       5.52       \\
                        &                          & 2015.12                        & 7.40    $\pm$ 0.74  & 0.35   $\pm$ 0.01    & $-$0.14   $\pm$ 0.01   & 0.06   $\pm$    0.001  & 252.13      $\pm$       25.21      \\
                        &                          & 2015.28                        & 8.98    $\pm$ 0.90  & $-$0.39  $\pm$ 0.02    & 0.16    $\pm$ 0.02   & 0.12   $\pm$    0.004  & 76.49       $\pm$       7.67       \\
                        &                          & 2015.36                        & 9.34    $\pm$ 0.93  & $-$0.46  $\pm$ 0.02    & 0.14    $\pm$ 0.02   & 0.10    $\pm$    0.004  & 114.56      $\pm$       11.41      \\
     &                          & 2015.44                        & 7.59    $\pm$ 0.76  & $-$0.50   $\pm$ 0.02    & 0.18    $\pm$ 0.02   & 0.08   $\pm$    0.003  & 145.47      $\pm$       14.57      \\
\hline
  \multirow{4}{*}{J4}                       &                          & 2015.12                        & 2.97    $\pm$ 0.30  & 0.15   $\pm$ 0.03    & $-$0.07   $\pm$ 0.03   & 0.14   $\pm$    0.03   & 18.59       $\pm$       1.88       \\
                        &                          & 2015.28                        & 1.85    $\pm$ 0.19  & $-$0.19  $\pm$ 0.04    & 0.05    $\pm$ 0.04   & 0.18   $\pm$    0.01   & 7.00         $\pm$       0.72       \\
                        &                          & 2013.36                        & 1.30    $\pm$ 0.13  & $-$0.32  $\pm$ 0.04    & 0.06    $\pm$ 0.04   & 0.21   $\pm$    0.03   & 3.62        $\pm$       0.36       \\
   &                          & 2015.44                        & 3.83    $\pm$ 0.40  & $-$0.46  $\pm$ 0.04    & 0.11    $\pm$ 0.04   & 0.20    $\pm$    0.02   & 11.74       $\pm$       1.23       \\
\hline
\multirow{2}{*}{J5}                         &                          & 2015.36                        & 2.45    $\pm$ 0.25  & $-$0.11  $\pm$ 0.02    & $-$0.05   $\pm$ 0.02   & 0.11   $\pm$    0.01   & 24.84       $\pm$       2.53       \\
   &                          & 2015.44                        & 2.81    $\pm$ 0.28  & $-$0.16  $\pm$ 0.02    & 0.01    $\pm$ 0.02   & 0.09   $\pm$    0.01   & 42.55       $\pm$       4.24       \\
     &                          & 2015.58 & 2.65 $\pm$  0.27 &  $-$0.21  $\pm$   0.02    & 0.01    $\pm$     0.02 &   0.11   $\pm$      0.01 & 7.51 $\pm$  0.75\\
\hline
 \multirow{2}{*}{J6} 
  &                          & 2015.99                        &  2.60 $\pm$ 0.30 & $-$0.19 $\pm$ 0.05 & 0.05  $\pm$ 0.05 & 0.22 $\pm$ 0.02 & 6.54 $\pm$ 0.65\\
 &                          & 2016.08                        & 1.93    $\pm$ 0.19  & $-$0.22  $\pm$ 0.04    & 0.10     $\pm$ 0.04   & 0.22   $\pm$    0.03   & 5.12        $\pm$       0.51       \\
\hline
J7                      &                          & 2016.31                        & 2.10     $\pm$ 0.21  & $-$0.16  $\pm$ 0.04    & 0.07    $\pm$ 0.04   & 0.19   $\pm$    0.02   & 7.10        $\pm$       0.72       \\
\hline
 \multirow{5}{*}{J8}                       &                          & 2016.44                        & 2.26    $\pm$ 0.22  & $-$0.14  $\pm$ 0.01    & 0.02    $\pm$ 0.01   & 0.07   $\pm$   0.01   & 34.22       $\pm$       3.33       \\
                        &                          & 2016.68                        & 2.84    $\pm$ 0.28  & $-$0.21  $\pm$ 0.02    & 0.06    $\pm$ 0.02   & 0.09   $\pm$    0.01   & 43.01       $\pm$       4.24       \\
                        &                          & 2016.80                         & 1.82    $\pm$ 0.18  & $-$0.20   $\pm$ 0.03    & 0.04    $\pm$ 0.03   & 0.15   $\pm$    0.01   & 9.92        $\pm$       0.98       \\
                        &                          & 2016.90                         & 1.41    $\pm$ 0.14  & $-$0.23  $\pm$ 0.04    & 0.02    $\pm$ 0.04   & 0.20   $\pm$    0.04   & 4.32        $\pm$       0.43       \\
    &                          & 2016.97                        & 1.70    $\pm$ 0.17  & $-$0.32  $\pm$ 0.04    & 0.09    $\pm$ 0.04   & 0.22   $\pm$    0.03   & 4.31        $\pm$       0.43       \\
                            &                          & 2017.09                        & 0.70   $\pm$ 0.10  & $-$0.38  $\pm$ 0.03    & 0.13    $\pm$ 0.03   & 0.13   $\pm$    0.02   & 77.66       $\pm$       7.98       \\
                        &                          & 2017.29                        & 0.30    $\pm$ 0.03  & $-$0.33  $\pm$ 0.02    & 0.24    $\pm$ 0.02   & 0.16   $\pm$    0.03   & 39.05       $\pm$       3.93
                        \\
\hline
\multirow{4}{*}{J9}                        &                          & 2016.90                        & 1.17    $\pm$ 0.12  & $-$0.03  $\pm$ 0.01    & $-$0.04  $\pm$  0.01   & 0.07   $\pm$    0.01   & 29.29       $\pm$       3.0        \\
                        &                          & 2016.97                        & 1.11    $\pm$ 0.11  & $-$0.12  $\pm$ 0.01    & $-$0.04   $\pm$ 0.01   & 0.11   $\pm$    0.02   & 5.32        $\pm$       0.53       \\
                        &                          & 2017.09                        & 2.01    $\pm$ 0.20  & $-$0.15  $\pm$ 0.03    & 0.001   $\pm$ 0.03   & 0.17   $\pm$    0.02   & 8.53        $\pm$       0.85       \\
    &                          & 2017.29                        & 0.94    $\pm$ 0.10  & $-$0.18  $\pm$ 0.03    & $-$0.03   $\pm$ 0.03   & 0.15   $\pm$    0.02   & 5.12        $\pm$       0.55 \\     
\hline  
\end{tabular}
\tablefoot{Columns from left to right: (1) component ID, (2) component ID in region C, (3) observed epoch, (4) flux density, (5) relative right ascension, (6) relative declination, (7) component size, and (8) brightness temperature.}
\end{table*}
\label{table:knots43b}

\begin{table*}
\caption{Model-fitting parameters at 86\,GHz.}
\label{table:kinem86}
\centering
\begin{tabular}{cccccccc}
\hline\hline  
Knot                   & Region C             & Time                           & S    & RA    & DEC   & FWHM   & $T_b$      \\
                       &                      & (Years)                        & (Jy) & (mas) & (mas) & (mas) & (10$^{\rm{10}}$K) \\
(1) & (2) & (3) & (4) & (5) & (6) & (7) & (8) \\ \hline
\multirow{8}{*}{Core}  &                      & 2013.73                        & 5.01   $\pm$   0.50   & $-$     & $-$            & 0.03    $\pm$    0.01   & 170.70       $\pm$       17.04     \\
                       &                      & 2014.39                        & 5.29    $\pm$  0.52   &  $-$      &    $-$      & 0.04    $\pm$    0.01   & 101.39       $\pm$       9.97      \\
                       &                      & 2014.72                        & 5.05    $\pm$  0.50   &    $-$      &    $-$      & 0.03    $\pm$    0.01   & 172.06       $\pm$       4.26     \\
                       &                      & 2015.37                        & 1.03    $\pm$  0.10   & $-$           & $-$         & 0.04    $\pm$    0.01   & 19.74        $\pm$       1.92      \\
                       &                      & 2015.72                        & 1.97    $\pm$  0.20   & $-$          & $-$             & 0.03    $\pm$    0.01   & 67.12        $\pm$       0.36      \\
                       &                      & 2016.38                        & 0.94    $\pm$  0.10    & $-$            & $-$             & 0.03    $\pm$    0.01   & 32.03        $\pm$       0.21      \\
                       &                      & 2016.74                        & 3.98    $\pm$  0.40   & $-$            & $-$              & 0.03   $\pm$   0.01   & 135.61         $\pm$       13.63      \\
                       &                      & 2017.24                        & 4.01   $\pm$   0.40   & $-$           & $-$             & 0.03   $\pm$    0.01   & 136.63       $\pm$       13.63      \\
\hline
    & C                        & 2013.73                        & 0.22    $\pm$   0.04   & $-$0.42    $\pm$   0.03   & 0.01     $\pm$   0.03  & 0.16    $\pm$    0.07   & 0.26       $\pm$       0.03       \\
\hline
            & \multirow{3}{*}{Ca} & 2015.72   & 2.81    $\pm$   0.30   & $-$0.57    $\pm$   0.02   & 0.25     $\pm$   0.02   & 0.11    $\pm$    0.01   & 7.12         $\pm$   0.76       \\
                       
                       &                      & 2016.38 & 0.70     $\pm$   0.10   & $-$0.55    $\pm$   0.02   & 0.20    $\pm$   0.02   & 0.15    $\pm$    0.03   & 0.95         $\pm$       0.14       \\
                       
                       &   & 2016.74                        & 0.40     $\pm$   0.04   & $-$0.45    $\pm$   0.04   & 0.21     $\pm$   0.04   & 0.18    $\pm$    0.05   & 0.38         $\pm$       0.04       \\
                       
\hline
                       &     \multirow{4}{*}{Cb} & 2015.72                        & 4.48    $\pm$   0.40   & $-$0.62    $\pm$   0.02   & 0.17     $\pm$  0.02   &  0.03   $\pm$    0.01   & 152.64        $\pm$       2.73       \\
                       &                      & 2016.38 & 0.70    $\pm$   0.07   & $-$0.58    $\pm$   0.01   & $-$0.01     $\pm$   0.01   & 0.08    $\pm$    0.02   & 3.35         $\pm$       0.38       \\
                       &                      & 2016.74                        & 2.05    $\pm$   0.20   & $-$0.59    $\pm$   0.03   & 0.13     $\pm$   0.03   & 0.13    $\pm$    0.02   & 3.72        $\pm$       0.36       \\
                       &                     & 2017.24                        & 0.27    $\pm$   0.03   & $-$0.54     $\pm$   0.02   & 0.02    $\pm$   0.02   & 0.09    $\pm$    0.04   & 1.02         $\pm$       0.11       \\
\hline
                       & \multirow{4}{*}{Cc}  & 2015.72                        & 0.85    $\pm$   0.08   & $-$0.61    $\pm$   0.01   & 0.05     $\pm$   0.02   & 0.13    $\pm$    0.02   & 1.54       $\pm$       0.82     \\
                       &                      & 2016.38 & 1.01    $\pm$   0.10   & $-$0.61    $\pm$   0.01   & 0.09     $\pm$   0.01   & 0.05    $\pm$    0.01   & 12.39        $\pm$       1.23      \\
                       &                      & 2016.74                        & 0.93    $\pm$   0.10   & $-$0.60     $\pm$   0.02   & $-$0.03    $\pm$   0.02   & 0.09    $\pm$    0.05   & 3.52         $\pm$       0.38       \\
                       &  & 2017.24                        & 1.47    $\pm$   0.20   & $-$0.49    $\pm$  0.03   & 0.13     $\pm$  0.03   & 0.14    $\pm$    0.04   & 2.30         $\pm$       0.31       \\
\hline
 \multirow{3}{*}{M}    &                      & 2014.39                        & 6.17    $\pm$   0.60   & $-$0.07    $\pm$   0.01   & 0.08     $\pm$   0.01   & 0.04    $\pm$    0.01   & 98.61       $\pm$       11.50      \\
                       &                      & 2014.72                        & 6.70    $\pm$   0.67   & $-$0.19    $\pm$   0.01   & 0.09     $\pm$   0.01   & 0.05    $\pm$    0.01   & 82.18        $\pm$       8.22     \\
    &                      & 2015.37                                               & 3.60    $\pm$   0.36   & $-$0.50     $\pm$   0.02   & 0.19     $\pm$   0.02   & 0.08    $\pm$   0.02   & 17.25        $\pm$       1.72       \\
\hline
 \multirow{2}{*}{J4}   &                      & 2014.72                        & 2.85    $\pm$   0.30    & $-$0.08   $\pm$ 0.01   & 0.01    $\pm$ 0.01   & 0.07    $\pm$    0.01   & 17.84        $\pm$       1.88       \\
   &                      & 2015.37                        & 0.84   $\pm$ 0.10    & $-$0.34   $\pm$ 0.03   & 0.10    $\pm$ 0.03   & 0.15    $\pm$    0.03   & 1.14         $\pm$       0.14       \\
\hline
\multirow{2}{*}{J5}      &                      & 2015.37                        & 0.95   $\pm$ 0.10  &   $-$0.15   $\pm$ 0.02   & 0.00   $\pm$ 0.02   & 0.12    $\pm$    0.03   & 2.02         $\pm$       0.21       \\
   &                      & 2015.72                        & 1.19   $\pm$ 0.10    & $-$0.33   $\pm$ 0.04   & 0.04    $\pm$ 0.04   & 0.18    $\pm$    0.04   & 1.13         $\pm$       0.10       \\
\hline
\multirow{2}{*}{J6}  &                      & 2015.72                        & 1.22   $\pm$ 0.12   & $-$0.08   $\pm$ 0.03   & $-$0.01   $\pm$ 0.03   & 0.15    $\pm$    0.03   & 1.66         $\pm$       0.16       \\
   &                      & 2016.38                        & 0.20    $\pm$ 0.02   & $-$0.38   $\pm$ 0.04   & 0.14    $\pm$ 0.04   & 0.12    $\pm$    0.05   & 0.43         $\pm$       0.04       \\
\hline
J7                     &                 & 2016.38                        & 0.50    $\pm$ 0.05    & $-$0.16   $\pm$ 0.01   & 0.04    $\pm$ 0.01   & 0.08    $\pm$    0.02   & 2.40         $\pm$       0.23       \\
\hline
\multirow{3}{*}{J8}  &                      & 2016.38                        & 0.14   $\pm$ 0.01   & $-$0.05   $\pm$ 0.01   & $-$0.02   $\pm$ 0.01   & 0.02    $\pm$    0.02   & 10.73        $\pm$       0.77     \\
   &                      & 2016.74                        & 0.80    $\pm$ 0.10   & $-$0.21   $\pm$ 0.01   & 0.03    $\pm$ 0.01   & 0.08    $\pm$    0.04   & 3.83         $\pm$       0.48       \\
      &                      & 2017.24                        &  0.22 $\pm$ 0.02 &  -0.30 $\pm$     0.02 & 0.16 $\pm$       0.02 & 0.08     $\pm$ 0.01 & 0.96 $\pm$ 0.10 \\
\hline
\multirow{2}{*}{J9}           &                  & 2016.74  & 0.54   $\pm$ 0.10    & $-$0.04   $\pm$ 0.01   & $-$0.02   $\pm$ 0.01   & 0.02    $\pm$    0.01   & 41.40        $\pm$       7.67       \\
   &                      & 2017.24                        & 0.62   $\pm$ 0.10    & $-$0.15   $\pm$ 0.01   & $-$0.02   $\pm$ 0.01   & 0.07    $\pm$    0.01   & 3.88         $\pm$       0.63       \\
\hline
J10                    &                      & 2017.24                        & 3.60   $\pm$ 0.40    & 0.05    $\pm$ 0.01   & $-$0.02   $\pm$ 0.01   & 0.04    $\pm$    0.01   & 69.04        $\pm$       6.90     \\
\hline  
\end{tabular}
\tablefoot{Columns from left to right: (1) component ID, (2) component ID in region C, (3) observed epoch, (4) flux density, (5) relative right ascension, (6) relative declination, (7) component size, and (8) brightness temperature.}
\label{table:knots86}
\end{table*}

\end{appendix}


\end{document}